\documentclass[12pt]{elsart}

\usepackage[dvips]{epsfig}
\usepackage{float}

\usepackage{bm}
\usepackage{amsmath}
\usepackage{amssymb}
\usepackage{latexsym}
\begin{document}

%
%
\hyphenation{brem-sstrah-lung proc-ess}
\hyphenation{quark quarks}
\hyphenation{hard}
\newcommand {\Et}       {{\rm E}_{\scriptscriptstyle\rm T}}
\newcommand {\Etmiss}   {{\not}{{\rm E}_{\scriptscriptstyle\rm T}}}
\newcommand {\Dslash}   {{\not}{{\rm D}}}
\newcommand {\Etcone}   {{\rm E}_{\scriptscriptstyle\rm T}^{cone}}
\newcommand {\Pt}       {{\rm P}_{\scriptscriptstyle\rm T}}
\newcommand {\Mtran}    {{\rm M}_{\scriptscriptstyle\rm T}}
\newcommand {\Mtop}     {{\rm M}_{\scriptscriptstyle\rm TOP}}
\newcommand {\Ete}      {\Et^{\rm e}}
\newcommand {\Ptmin}    {\Pt^{\rm min}}
\newcommand {\Ptmu}     {\Pt^{\mu}}
\newcommand {\Degs}     {^\circ}
\newcommand {\Stw}      {\sin^{2}\theta_{W}}
\newcommand {\Amp}      {{\cal A}}
\newcommand {\Lk}       {{\cal L}}
\newcommand {\Prob}     {{\cal P}}
\newcommand {\Roots}    {{\sqrt{\rm s}}}
\newcommand {\Deltar}   {\sqrt{\Delta\eta^{2}+\Delta\phi^{2}}}
\newcommand {\Meta}     {{|\eta|}}
\newcommand {\Eplus}    {{\epsilon_{\rm \scriptscriptstyle POSITIVO}}}
\newcommand {\Eminus}   {{\epsilon_{\rm \scriptscriptstyle NEGATIVO}}}
\newcommand {\Qbar}     {{\bar {\rm q}}}
\newcommand {\Emu}      {{{\rm e}\mu}}
\newcommand {\W}[1]     {{W^{#1}}}
\newcommand {\Z}[1]     {{Z^{#1}}}
\newcommand {\Ks}       {K_{s}^{0}}
\newcommand {\Pich}     {\pi^{\pm}}
\newcommand {\Piz}      {\Pi^{0}}
\newcommand {\Ppbar}    {p\bar{p}}
\newcommand {\Qqbar}    {q\bar{q}}
\newcommand {\Ttbar}    {t\bar{t}}
\newcommand {\Udbar}    {u\bar{d}}
\newcommand {\Bbbar}    {b\bar{b}}
\newcommand {\Ccbar}    {c\bar{c}}
\newcommand {\Epem}     {{\rm e^{+}e^{-}}}
\newcommand {\Mpmm}     {{\mu^{+}\mu^{-}}}
\newcommand {\Tptm}     {{\tau^{+}\tau^{-}}}
\newcommand {\Wpthree}  {{{\rm W} \; + \geq 3}}
\newcommand {\Wpfour}   {{{\rm W} \; + \geq 4}}
%
%
\newcommand {\Wenu}     {{\W{}\rightarrow{\rm e}\nu}}
\newcommand {\Wmunu}    {{\W{}\rightarrow\mu\nu}}
\newcommand {\Wlep}     {{\W{}\rightarrow{\rm l}\nu}}
\newcommand {\Ppbb}     {{\Ppbar\rightarrow\Bbbar}}
\newcommand {\Bele}     {{{\rm b}\rightarrow{\rm c}{\rm e}\nu_{\rm e}}}
\newcommand {\Blnu}     {{{\rm b}\rightarrow{\rm c}{\rm l}\nu_{\rm l}}}
\newcommand {\Ztau}     {{\Z{}\rightarrow\tau^{+}\tau^{-}}}
\newcommand {\Zmumu}    {{\Z{}\rightarrow\mu^{+}\mu^{-}}}
\newcommand {\Zee}      {{\Z{}\rightarrow{\rm e}^{+}{\rm e}^{-}}}
\newcommand {\Temux}    {{\Ttbar \rightarrow \Emu + {\rm X}}}
\newcommand {\Drellyan} {{{\rm q}\Qbar\rightarrow\left(\gamma,\Z{0}\right)
                        \rightarrow{\rm l}^{+}{\rm l}^{-}}}
\newcommand {\SMgauge}  {{SU(3) \otimes SU(2)_L \otimes U(1) }}
\newcommand {\EWgauge}  {{SU(2)_L \otimes U(1) }}
\newcommand{\LL}{\ell^+ \ell^-}
\newcommand{\EE}{\rm e^+ e^-}
\newcommand{\MM}{\mu^+ \mu^-}
\newcommand{\TT}{\rm t \bar{t}}
\newcommand{\QQ}{\rm q \bar{q}}
\newcommand{\BB}{\rm b \bar{b}}
\newcommand{\CC}{\rm c \bar{c}}
\newcommand{\NN}{\nu \bar{\nu}}
\newcommand{\WW}{\rm W^+ W^-}

%
%
\newcommand {\Gev}      {{{\rm GeV/c}^{2}}}
\newcommand {\Mum}      {{\mu{\rm m}}}
\newcommand {\Mus}      {{\mu{\rm s}}}
\newcommand {\Cms}      {{{\rm cm}^{-2}{\rm s}^{-1}}}
\newcommand {\Imb}      {{\mu{\rm b}^{-1}}}
\newcommand {\Inb}      {{{\rm nb}^{-1}}}
\newcommand {\Ipb}      {{{\rm pb}^{-1}}}
\newcommand {\Ifb}      {{{\rm fb}^{-1}}}
%
%
\newcommand {\etaltri}     {{\textit{ et al.}}}
\newcommand {\phm}[1]   { Phil. Mag. {\bf #1}}
\newcommand {\nat}[1]   { Nature {\bf #1}}
\newcommand {\prl}[1]   { Phys. Rev. Lett {\bf #1}}
\newcommand {\prev}[1]  { Phys. Rev. {\bf #1}}
\newcommand {\prd}[1]   { Phys. Rev. D {\bf #1}}
\newcommand {\zs}[1]    { Z. Phys. {\bf #1}}
\newcommand {\sov}[1]   { Sov. J. Nucl. Phys. {\bf #1}}
\newcommand {\ncim}[1]  { Nuovo Cim. {\bf #1}}
\newcommand {\plet}[1]  { Phys. Lett. {\bf #1}}
\newcommand {\prep}[1]  { Phys. Rep. {\bf #1}}
\newcommand {\rmp}[1]   { Rev. Mod. Phys. {\bf #1}}
\newcommand {\nphy}[1]  { Nucl. Phys. {\bf #1}}
\newcommand {\nim}[1]   { Nucl. Instrumen. Meth. {\bf #1}}
%
%
\newcommand {\pawfig}[1]        {\includegraphics[width=6.4cm]{#1}}
\newcommand {\doublepawfig}[2]  {
                                  \begin{tabular}{cc}
                \includegraphics[height=6cm,clip]{#1} &
                \includegraphics[height=6cm,clip]{#2}
                                  \end{tabular}                         }
%
%
\newcommand {\et}       {$\Et\;$}
\newcommand {\etmiss}   {$\Etmiss\;$}
\newcommand {\etcone}   {$\Etcone\;$}
\newcommand {\pt}       {$\Pt\;$}
\newcommand {\mtran}    {$\Mtran\;$}
\newcommand {\mtop}     {$\Mtop\;$}
\newcommand {\ete}      {$\Ete\;$}
\newcommand {\ptmin}    {$\Ptmin\;$}
\newcommand {\ptmu}     {$\Ptmu\;$}
\newcommand {\degs}     {$\Degs\;$}
\newcommand {\stw}      {$\Stw\;$}
\newcommand {\lk}       {$\Lk\;$}
\newcommand {\roots}    {$\Roots\;$}
\newcommand {\deltar}   {$\Deltar\;$}
\newcommand {\meta}     {$\Meta\;$}
\newcommand {\eplus}    {$\Eplus\;$}
\newcommand {\eminus}   {$\Eminus\;$}
\newcommand {\qbar}     {$\Qbar\;$}
\newcommand {\emu}      {$\Emu\;$}
\newcommand {\w}[1]     {$\W{#1}\;$}
\newcommand {\z}[1]     {$\Z{#1}\;$}
\newcommand {\ks}       {$\Ks\;$}
\newcommand {\pich}     {$\Pich\;$}
\newcommand {\piz}      {$\Piz\;$}
\newcommand {\ppbar}    {$\Ppbar\;$}
\newcommand {\qqbar}    {$\Qqbar\;$}
\newcommand {\ttbar}    {$\Ttbar\;$}
\newcommand {\bbbar}    {$\Bbbar\;$}
\newcommand {\ccbar}    {$\Ccbar\;$}
\newcommand {\epem}     {$\Epem\;$}
\newcommand {\mpmm}     {$\Mpmm\;$}
\newcommand {\tptm}     {$\Tptm\;$}
\newcommand {\wenu}     {$\Wenu\;$}
\newcommand {\wmunu}    {$\Wmunu\;$}
\newcommand {\wlep}     {$\Wlep\;$}
\newcommand {\ppbb}     {$\Ppbb\;$}
\newcommand {\bele}     {$\Bele\;$}
\newcommand {\blnu}     {$\Blnu\;$}
\newcommand {\ztau}     {$\Ztau\;$}
\newcommand {\zmumu}    {$\Zmumu\;$}
\newcommand {\zee}      {$\Zee\;$}
\newcommand {\temux}    {$\Temux\;$}
\newcommand {\drellyan} {$\Drellyan\;$}
\newcommand {\gev}      {$\Gev\;$}
\newcommand {\mum}      {$\Mum\;$}
\newcommand {\mus}      {$\Mus\;$}
\newcommand {\cms}      {$\Cms\;$}
\newcommand {\imb}      {$\Imb\;$}
\newcommand {\inb}      {$\Inb\;$}
\newcommand {\ipb}      {$\Ipb\;$}
\newcommand {\wpthree}  {$\Wpthree\;$}
\newcommand {\wpfour}   {$\Wpfour\;$}
%
%
\newcommand {\ii}       {\`{\i} }
%
%

%
\pagestyle{plain}
\begin{frontmatter}
\title{\Large{\bf{Bose-Einstein Correlations in charged current muon-neutrino interactions in the NOMAD experiment at CERN}}}
\collab{NOMAD Collaboration}
\author[Paris]             {P.~Astier}
\author[CERN]              {D.~Autiero}
\author[Saclay]            {A.~Baldisseri}
\author[Padova]            {M.~Baldo-Ceolin}
\author[Paris]             {M.~Banner}
\author[LAPP]              {G.~Bassompierre}
\author[Lausanne]          {K.~Benslama} 
\author[Saclay]            {N.~Besson}
\author[CERN,Lausanne]     {I.~Bird}
\author[Johns Hopkins]     {B.~Blumenfeld}
% ...

\address[LAPP]           {LAPP, Annecy, France}                               
\address[Johns Hopkins]  {Johns Hopkins Univ., Baltimore, MD, USA}            
% ...
\address[CERN]           {CERN, Geneva, Switzerland}                          
\address[Lausanne]       {University of Lausanne, Lausanne, Switzerland}      
% ...
\address[Padova]         {Univ. of Padova and INFN, Padova, Italy}            
\address[Paris]          {LPNHE, Univ. of Paris VI and VII, Paris, France}    
% ...
\address[Saclay]         {DAPNIA, CEA Saclay, France}                         
% ...

\begin{abstract}
Bose-Einstein Correlations in one and two dimensions have been studied, with high
statistics, in charged current muon-neutrino interaction events collected with the
NOMAD detector at CERN. In one dimension the Bose-Einstein effect has been analyzed with
the Goldhaber and the Kopylov-Podgoretskii phenomenological parametrizations.
The Goldhaber parametrization gives  the radius of the pion emission region 
$R_G = 1.01\pm 0.05$(stat)$^{+0.09}_{-0.06}$(sys) fm  and for the chaoticity parameter the value $\lambda = 0.40\pm 0.03$(stat)$^{+0.01}_{-0.06}$(sys).  Using the
Kopylov-Podgoretskii parametrization yields  $R_{KP} = 2.07\pm 0.04$(stat)$^{+0.01}_{-0.14}$(sys) fm and $\lambda_{KP} = 0.29\pm 0.06$(stat)$^{+0.01}_{-0.04}$(sys). Different parametrizations of the long-range correlations have been also studied.  The
two-dimensional shape of the source has been investigated in the longitudinal
co-moving frame.  A significant difference between the transverse and the longitudinal
dimensions is observed.  The high statistics of the collected sample allowed the study of the
Bose-Einstein correlations as a function of rapidity, charged particle multiplicity and
hadronic energy. A weak dependence of both radius and chaoticity on multiplicity and hadronic energy
is found.
\end{abstract}
\clearpage
\end{frontmatter}

\section{Introduction}
The quantum mechanical wave function of two identical bosons has to be symmetric
under particle exchange.  The symmetrization gives rise to an observable interference
pattern which enhances the number of identical bosons emitted close to one another
in phase space.  Such Bose-Einstein Correlations (BEC) were observed
for the first time in astronomical measurements of photon pairs emitted by stars
\cite{ref:twiss} and soon after for like-sign hadrons produced in $p\bar{p}$
annihilations \cite{ref:Goldhaber}. Since then BEC, were also measured in several other types of particle interactions
(for a review see \cite{ref:weiner}).
The shape of the BEC depends on the spatial and
temporal distributions of the boson source and on its degree of coherence. The theoretical aspects of the BEC were developed in the papers of Kopylov and Podgoretskii \cite{ref:Kopylov} and Cocconi \cite{ref:Cocconi}.
From these  studies it appears that the measurements of BEC may be important to gain an understanding on the dynamics of the
particle interactions yielding like-sign bosons in the final state.

Previous measurements of the BEC effects in neutrino interactions have been performed by the Big
European Bubble Chamber Collaboration (BEBC) \cite{ref:BEBC} including data
collected on a variety of targets by both BEBC at CERN and the 15-foot Bubble Chamber at Fermilab.
  Nevertheless, the number of events
globally collected by these experiments is still about one order of
magnitude smaller than the data set collected by NOMAD and used in
this paper.
\section{The phenomenology of BEC}
The  BEC effect can be parametrized in terms of the two particle
correlation function $R$ defined as:
\begin{equation}
    R(p_1,p_2) = D(p_1,p_2)/D_{0}(p_1,p_2)
\label{eq:ratio}
\end{equation}
where $p_{1,2}$ are the particle four-momenta, $D(p_{1},p_{2})$ is the measured
two-particle density and $D_{0}(p_{1},p_{2})$, the particle density in the absence
of BEC. $D_{0}(p_{1},p_{2})$ should include any other two-particle correlations such as those coming from phase space, long-range correlations, charge effects, etc. which in the ratio
should be divided out leaving only the BEC effects.
According to the Goldhaber parametrization \cite{ref:Goldhaber}, which assumes that the emitting sources of identical bosons are described by a spherical Gaussian density function,
BEC are usually parametrized as:

\begin{equation}
    R(Q) = 1 + \lambda \exp\big(-R^2_G Q^2\big)
\label{eq:erre}
\end{equation}
where $Q^2 = -(p_1 - p_2)^2 = M^2_{\pi\pi}-4m^2_{\pi}$, with
$M_{\pi\pi}$ the
invariant mass of the pion pair, $R_G$  the width of the Gaussian
distributed emitting source and $m_{\pi}$ the pion mass. The chaoticity (or
incoherence) parameter $\lambda$ measures the degree of
coherence in  pion production, i.e. the fraction of pairs of identical particles that undergo interference
($0 \leqslant \lambda \leqslant 1$).

The Kopylov-Podgoretskii (KP) parametrization \cite{ref:Kopylov} corresponds to a
radiating spherical surface of radius $R_{KP}$ with pointlike oscillators of
lifetime $\tau$:
\begin{equation}
    R(Q_t,Q_0) = 1 + \lambda\big[4J^2_1(Q_t R_{KP})/(Q_t R_{KP})^2\big]/\big[1 + (Q_0 \tau)^2\big]
\label{eq:kopylov2}
\end{equation}
where $J_1$ is the first-order Bessel function, $\vec{p} = \vec{p}_1 + \vec{p}_2$, $\vec{Q} = \vec{p}_1 - \vec{p}_2$,
$Q_0 = \left| E_1 - E_2 \right|$, $Q_t = \left| \vec{Q} \times \vec{p} \right|/\left| \vec{p} \right|$.  This
parametrization is not Lorentz invariant and the variables are calculated in the
centre of mass of the hadronic final state.  It can be shown that at small
values of $Q_t$ and $Q_0$ the relation $R_{KP} \approx 2R_G$ is expected \cite{ref:BEBC}.

\begin{figure}[tb]
\begin{center}
\centerline{\input{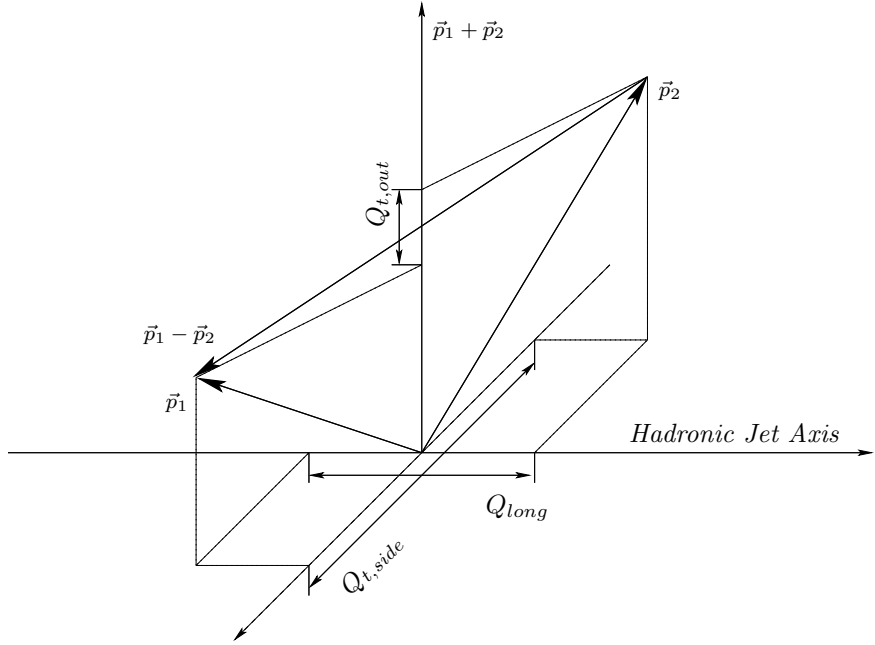}}
\caption{\small{\em The definition of the LCMS system.}}
\label{fig:LCMS}
\end{center}
\end{figure}

The shape of the hadronic source can be measured by studying BEC as a
function of the components of the vector $\vec{Q}$.  It
is convenient to perform this study in the so-called longitudinal centre of mass
system (LCMS).  This reference system is defined for every particle pair as that
where $\vec{p} = \vec{p}_1 + \vec{p}_2$ is perpendicular to the axis defined by
the hadronic jet direction (see fig. \ref{fig:LCMS}).  With this choice, possible
effects caused by the Lorentz boost are avoided.  In the LCMS, $\vec{Q}$ is decomposed
into the following components: $Q_{long}$, parallel to the hadronic jet axis;
$Q_{t,out}$, collinear with $\vec{p}$ and the
complementary $Q_{t,side}$, perpendicular to both $Q_{long}$ and $Q_{t,out}$.  In
this analysis we use the 
longitudinal component $Q_{||} = Q_{long}$ and the perpendicular component
$Q_{\bot} = \sqrt{Q^2_{t,out}+Q^2_{t,side}}$ (see fig.\ref{fig:LCMS}).  The
parametrization of the correlation is then performed  separately
for the longitudinal $Q_{\|}$ and transverse $Q_{\bot}$ components as suggested in ref \cite{ref:Goldhaber}:
\begin{equation}
    R(Q_{\|},Q_{\bot}) = 1 + \lambda\exp\big(-Q^2_{\|} R^2_{\|} - Q^2_{\bot} R^2_{\bot}\big)
\label{eq:LCMSeq}
\end{equation}
the longitudinal and transverse dimensions of the hadron source being represented
by $R_{\|}$ and $R_{\bot}$, respectively.
\section{Experimental procedure}
\subsection{The NOMAD experiment}
The main goal of the NOMAD experiment \cite{ref:NOMADET} was the search for
$\nu_{\mu}\rightarrow\nu_{\tau}$ oscillations in a wide-band neutrino beam
from the CERN SPS.  The full data sample, corresponding
to about 1.3 million $\nu_{\mu}$ charged-current (CC) interactions collected in four years of
data taking (1995--1998) in the detector fiducial volume, is used in the
present analysis.  The data are compared to the results of a Monte Carlo
simulation based on modified versions of the  LEPTO 6.1 \cite{ref:LEPTO} and the
JETSET 7.4 \cite{ref:JETSET} generators for neutrino interactions and on a
GEANT 3.21 \cite{ref:GEANT} based program for the detector response.  BEC effects
are not included in the Monte Carlo. For the analysis reported below we have
used a Monte Carlo sample of size comparable to the data.
\subsection{The NOMAD detector}
\begin{figure}[htb]
\begin{center}
\centerline{\epsfig{file=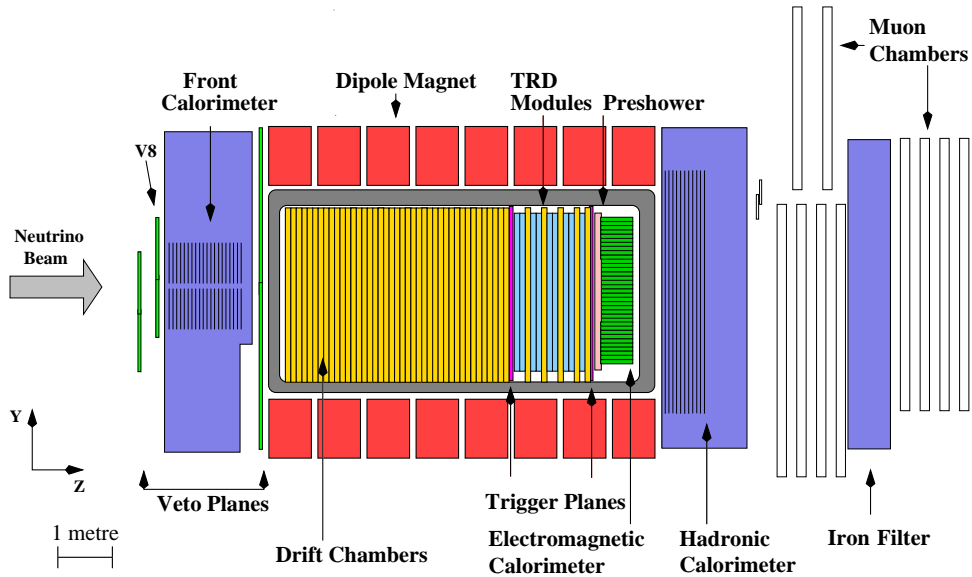,width=0.8\linewidth,angle=0}}
\caption{\small{\em NOMAD Apparatus.}}
\label{fig:nomadside}
\end{center}
\end{figure}
The tracking capabilities of the detector are essential for the study of BEC.
The NOMAD detector shown in fig. \ref{fig:nomadside} is especially
well suited for this.  It consists of an active target of 44 drift chambers,
with a total fiducial mass of 2.7 tons located in a 0.4 T dipole magnetic
field.  The drift chambers (DC) \cite{ref:DC}, made of low Z material (mainly
carbon), serve the dual role of a nearly isoscalar target for neutrino
interactions and of a tracking medium.  These drift chambers provide an
overall efficiency for charged particle reconstruction greater than 95\%
and a momentum resolution which can be parametrized as
$\frac{\sigma_{\left|\vec{p}\right|}}{\left|\vec{p}\right|}=\frac{0.05}{\sqrt{L}}\oplus\frac{0.008\left|\vec{p}\right|}{\sqrt{L^{5}}}$
where the track length $L$ is in metres and the track momentum $\left|\vec{p}\right|$ in GeV/c.  This
amounts to a resolution $\leqslant 3.5\%$ for $\left|\vec{p}\right| \leqslant 10$ GeV/c.  Reconstructed
tracks are used to determine the event topology (the assignment of tracks to
vertices) and to reconstruct the vertex position and the track parameters at each
vertex (primary, secondary, V{0}, etc\ldots).  A transition radiation detector (TRD)
\cite{ref:TRD} is used for electron identification.  The pion rejection achieved
for isolated tracks is $10^{3}$ with a 90\% electron identification efficiency.  A
lead-glass electromagnetic calorimeter (ECAL) \cite{ref:ECAL} located downstream of the
tracking region provides an energy resolution of $3.2\%/\sqrt{E \rm{[GeV]}}\oplus 1\%$
for electromagnetic showers and is essential to measure the total energy flow
in neutrino interactions.  In addition an iron absorber and a set of drift
chambers located after the electromagnetic calorimeter are used for muon
identification, providing a muon detection efficiency of 97\% for momenta
greater than 5 GeV/c.
\subsection{Event selection}
The identification of $\nu_{\mu}$CC events requires the presence of a primary
negative muon in the final state, i.e. a track segment in the muon detector
matched to a track reconstructed in the drift chambers.  The muon momentum and
its transverse component (relative to the neutrino beam) are required to be
greater than 5 GeV/c and 0.5 GeV/c, respectively.  Preliminary cuts
are applied to ensure good quality event reconstruction\footnote{additional details on the analysis
can be found in \cite{ref:theses}.}:
\begin{itemize}
\item Number of primary charged tracks (excluding the muon) $N_{ch} \geqslant 2$;
\item Muon energy $E_{\mu} \geqslant 5$ GeV and hadronic energy $E_{hadrons} \geqslant 5$ GeV;
\item Hadronic invariant mass $W \geqslant 2$ GeV (to reject quasi-elastic events and baryon resonance production);
\item Event vertex within the fiducial region of the DC target.
\end{itemize}
Tracks to be used for BEC are then selected using the following criteria:
\begin{itemize}
\item Only primary tracks are selected, i.e. either belonging or pointing to the primary
    vertex; in the latter
    case, to avoid potential dangerous contamination from photon conversions, the
    track first hit must occur no further than 15 cm downstream of the vertex along the detector axis (the z axis); 
\item A minimum momentum of the track is required: $\left|\vec{p}_{track}\right|
    \geqslant 100$ MeV/c;
\item A minimum number of hits is used to build the track: $N_{hits} \geqslant 12$;
\item A good momentum resolution is required: $\frac{\Delta \left|\vec{p}\right|}{\left|\vec{p}\right|} \leqslant 6\%$, where $\Delta\left|\vec{p}\right|$ is the the uncertainty on the momentum of that track calculated by the reconstruction program;
\item The track should not be identified as an electron by the TRD and the ECAL;
\item The track should not be identified as a proton by the range-momentum correlation method
(see ref. \cite{ref:veltri} for details).
\end{itemize}
The total number of events selected by these cuts is  398K. These events contain  544K (++), 143K ($--$), and 852K ($+-$) pairs.

After the event and track quality cuts, we have performed a preliminary analysis of the simulated events  to assess the purity of the track identification. The tracks investigated are those obtained from the full Monte Carlo simulation. An appropriate  algorithm allowed an association between the reconstructed tracks and those generated at the primary vertex. We have found that the positive and negative samples of particles used for BEC studies contain respectively $\approx 61\%$ of
${\pi}^+$ and $\approx 77\%$ of ${\pi}^-$. The relative contributions of various positive and negative particles in NOMAD entering the correlation plots are listed in  
 table \ref{tab:percentage}.
\begin{table}[htb]
\begin{center}
\begin{tabular}{|c|c|}
\hline
Particle & Percentage \\
\hline
${\pi}^+$      & 61.2\% \\
$p$            & 18.4\% \\
$K^+$          & 5.6\%  \\
${\mu}^+$      & 0.1\%  \\
$e^+$          & 0.2\%  \\
Not recognized & 14.5\% \\
\hline
${\pi}^-$      & 77.2\% \\
$\bar{p}$      & 2.4\%  \\
$K^-$          & 7.8\%  \\
$e^-$          & 0.3\%  \\
Not recognized & 12.3\% \\
\hline
\end{tabular}
\caption{\small{\em Composition of the charged particle sample used in the analysis.}}
\label{tab:percentage}
\end{center}
\end{table}
The tracks labeled ``not recognized'' are those for which no association with a generated primary track was found. These tracks
are produced by secondary interactions or photon-conversions.  As we
can see from table \ref{tab:percentage} the negative tracks exhibit a better pion purity
than the positive ones.  Contaminations from electrons and positrons which are mainly present in the "not recognized" samples could be
dangerous since these particles come from photon conversions and therefore they can populate the
low Q region where BEC effects are expected.  This problem, together with
the effects caused by the proton and kaon contaminations, will be discussed in
section 4.

The Monte Carlo simulation was also used to verify the experimental resolution in the determination of the BEC parameters. BEC effects are expected to occur in any of the kinematical variables $Q$, $Q_t$,
$Q_0$, $Q_{\|}$, $Q_{\bot}$ at small values of these parameters ($\leqslant 0.2$ GeV) and we have verified  that the  resolution in any of these variables is
$\leqslant 0.02$ GeV.  The BEC parameters are obtained by fits to the experimental distributions in the interval 0.0-1.5 GeV. This interval is large enough
 to study   possible long-range correlations as well.
\section{ The reference samples}
We have studied several alternatives for the choice of the reference sample $D_0(p_1,p_2)$ used in eq. \ref{eq:ratio}.  In principle the Monte Carlo events,
which do not contain BEC, would be good candidates. However the capability
of the Monte Carlo to accurately reproduce  the data (except for BEC) especially in the
tiny phase-space region where BEC are present is limited and other methods based
on the data themselves must be found.  Several methods have been used in previous experiments
to build the reference sample from the data (see for example \cite{ref:BEBC}). They are:
\begin{itemize}
\item The reference sample is formed of all unlike-sign pairs;
\item The reference sample is formed by building a so called ``mixed event'': a hadron from one event is combined with a hadron of the same charge, chosen at random from an another event that has approximately the same kinematical characteristics: total hadron momentum, hadron energy, charged multiplicity...
\item The reference sample is formed by pairing unlike-charge hadrons from the same event after the transverse momenta $\vec{p}_{t}$ (with respect to the current direction) have been  interchanged at random in the hadronic centre-of-mass system (c.m.s).
\end{itemize}
We have carefully tested the three methods with a full Monte Carlo simulation (discussed in more detail later in this section) which includes also the response of the detector. The Monte Carlo results reproduce correctly the inclusive particle distribution in neutrino interactions, but not the correlations among particles. This Monte Carlo is therefore adequate to study the bias introduced by the reference sample. In fact, in the absence of BEC,  the distributions in R should be flat or, in any case, have no structure at small Q ($\leqslant 0.2$ GeV) which would distort the study of R. We have found that none of the three reference samples completely fulfills this requirement, however the unlike-sign one was eventually found to be the most adequate in the BEC region.
Moreover
this reference sample has been used by the great majority of previous
BEC studies.

BEC effects  are  then investigated by looking in the data at the following
ratio:
\begin{equation}
    R(Q) = \frac{\mbox{``like-sign''pion-pairs}}{\mbox{``unlike-sign''pion-pairs}} = \frac{N_{++}(Q)+N_{--}(Q)}{N_{+-}(Q)}
\label{eq:like/unlike}
\end{equation}
The Monte Carlo samples are used to estimate possible spurious BEC effects 
from non-pion contaminations present in the sample as discussed in the
following section.
\subsection{Systematic effects}
\begin{figure}[htb]
\begin{minipage}{.5\linewidth}
\begin{center}
\centerline{\epsfig{file=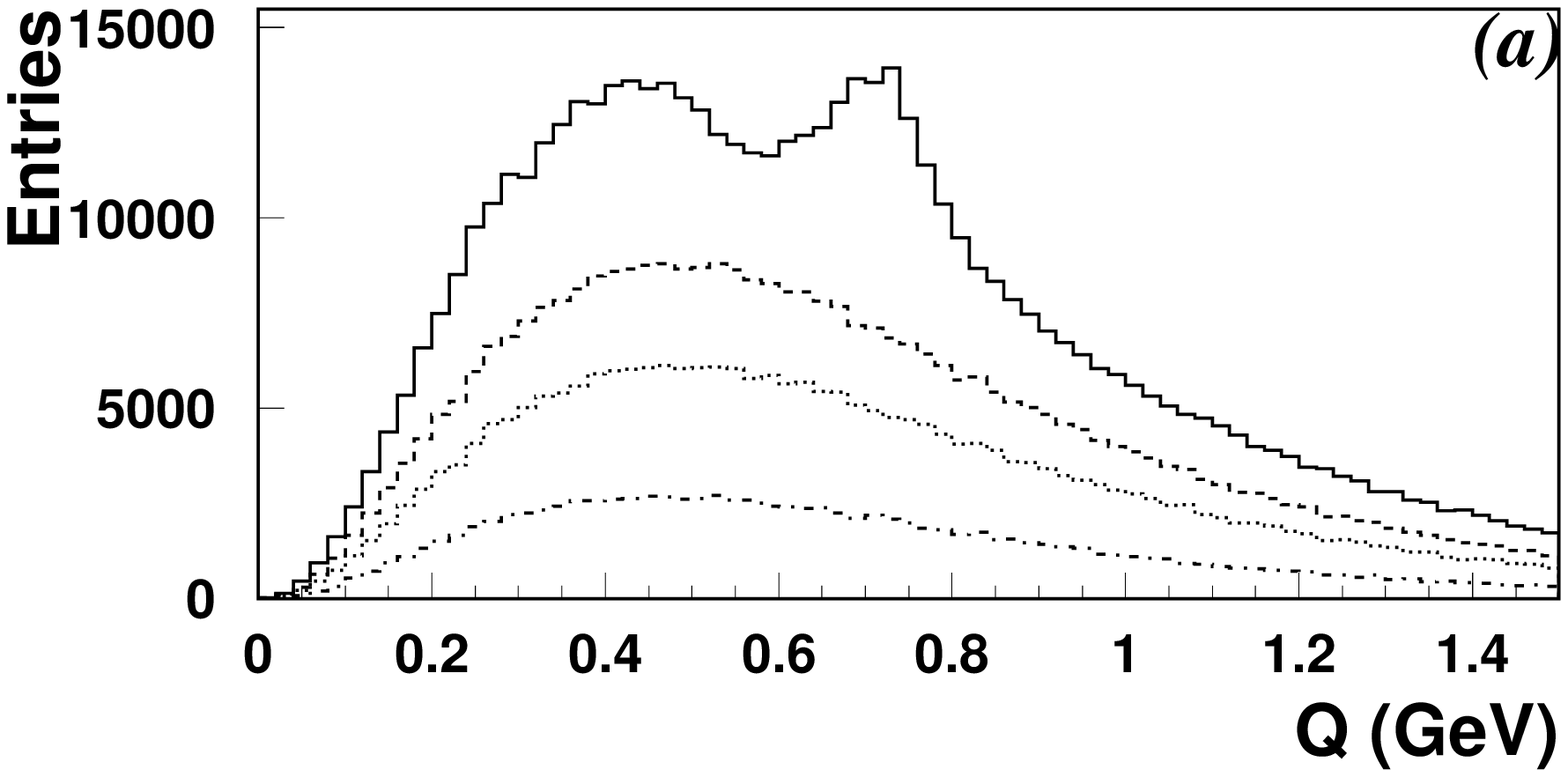,width=0.9\linewidth,angle=0}}
\end{center}
\end{minipage} \hfill
\begin{minipage}{.5\linewidth}
\begin{center}
\centerline{\epsfig{file=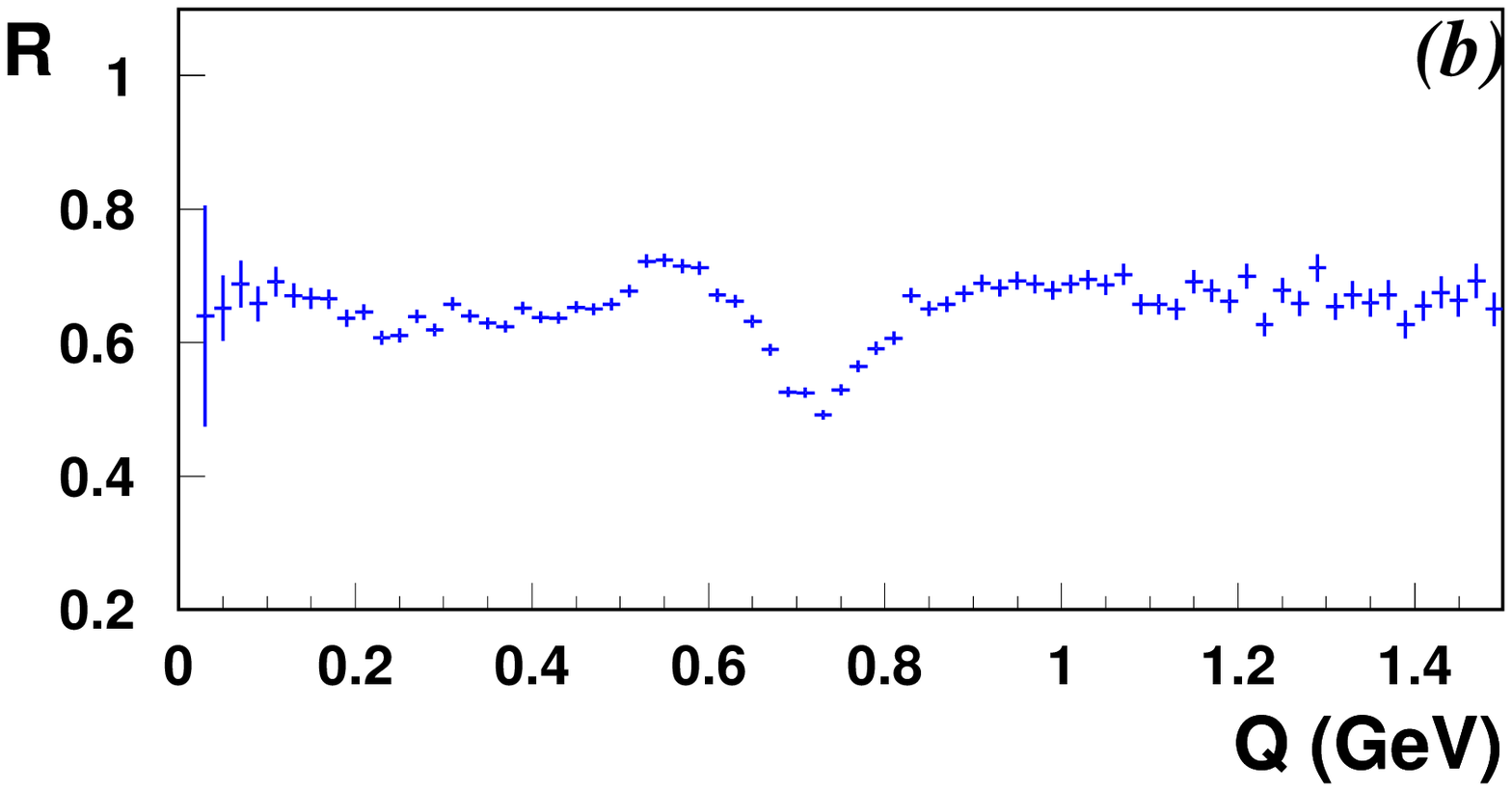,width=0.9\linewidth,angle=0}}
\end{center}
\end{minipage}
\caption{\small{\em (a), Q-distributions for Monte Carlo generated pion pairs; top to bottom:
unlike, like, (+ +), ($--$) pairs. (b),  the ratio like/unlike pairs, R(Q) of
eq. \ref{eq:like/unlike}, for a pure pion MC sample.}}
\label{fig:mcgen_pions}
\end{figure}

\begin{figure}[htb]
\begin{minipage}{.5\linewidth}
\begin{center}
\centerline{\epsfig{file=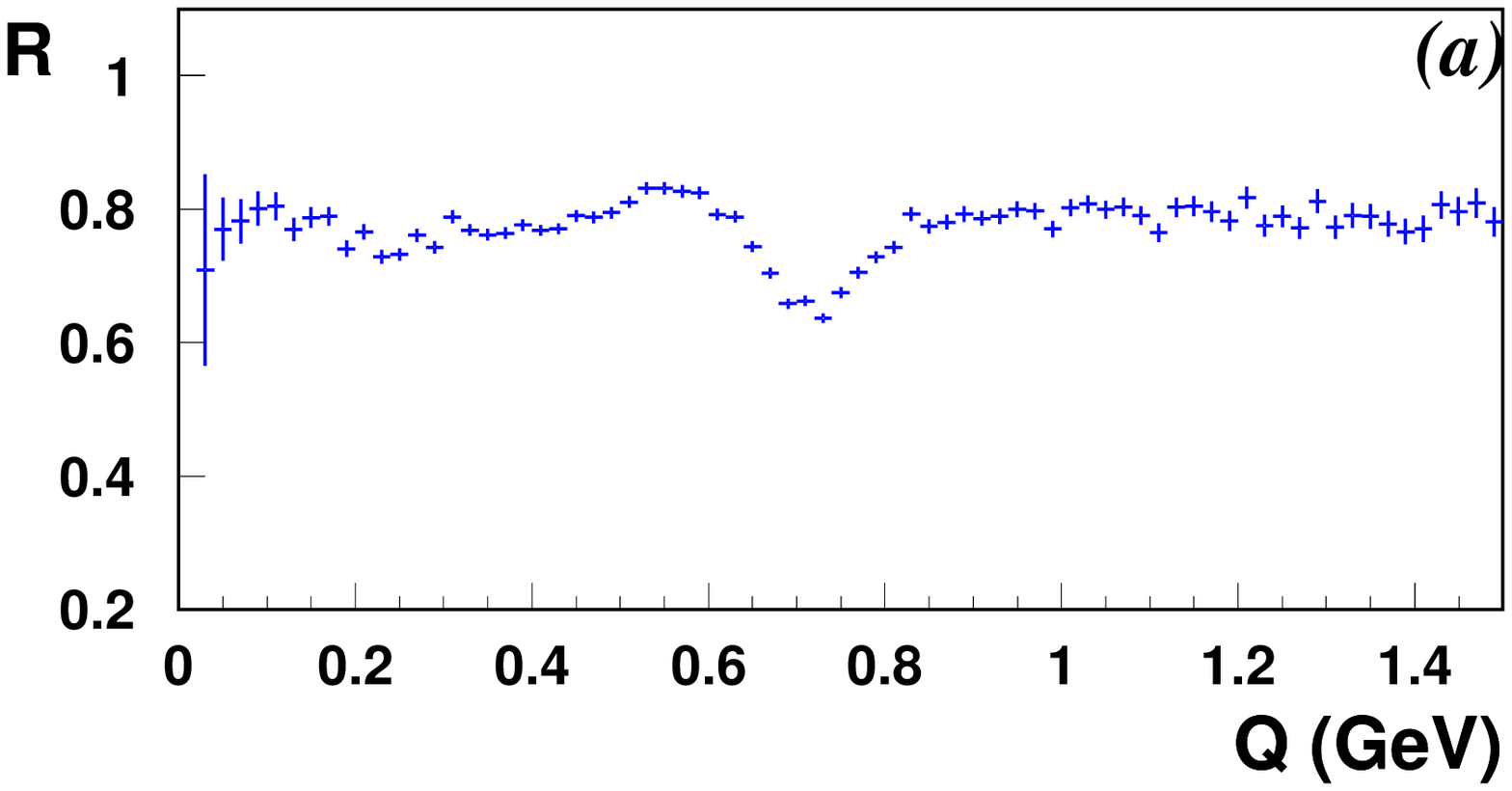,width=0.9\linewidth,angle=0}}
\end{center}
\end{minipage} \hfill
\begin{minipage}{.5\linewidth}
\begin{center}
\centerline{\epsfig{file=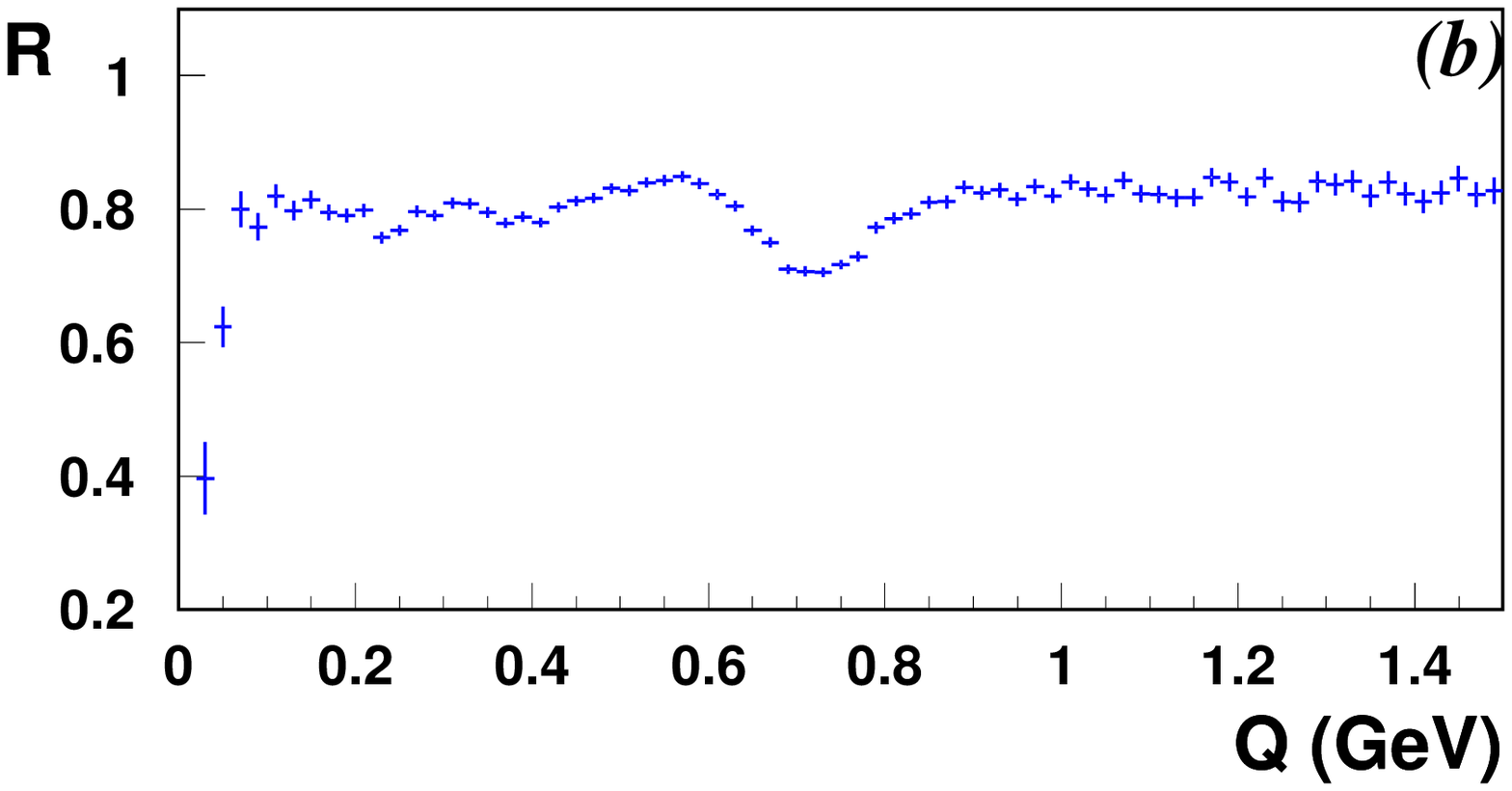,width=0.9\linewidth,angle=0}}
\end{center}
\end{minipage}
\caption{\small{\em (a), R(Q) for a Monte Carlo sample of pions, kaons and protons. (b), R(Q) for a Monte Carlo sample of all particle types. }}
\label{fig:mcgen_all}
\end{figure}

In the analysis, all
secondary charged particles have been assumed to be pions, unless identified as
muons, electrons or protons. However, as seen in section 3 for simulated events, there is a fraction of 39\% and 23\%, respectively
of these positive and negative particles, which are not pions. Here we want to study how the BEC could be changed by these misidentified tracks and by the use of the unlike-sign sample as a reference.  These effects could manifest themselves in 
three distinct ways:
\begin{itemize}
\item BEC for like-sign kaons and correlations for like-sign
fermions (electrons, protons).  The BEC for kaons were measured at LEP and 
showed  characteristics very similar to those of pions.  Fermion pairs, instead, could exhibit
an anticorrelation effect. However, the number of like-sign kaon and fermion pairs is very small and their contribution is negligible.
\item Pairs of like-sign, but not identical particles in the numerator of
eq. \ref{eq:like/unlike}, for example $K^{+}\pi^{+}$ pairs, have no BEC. This contribution
could bias the value of $\lambda$.
\item  Unlike-sign pairs in the denominator of eq. \ref{eq:like/unlike}
include contributions from  $K^0$ and resonances such as $\rho$, $\omega$, as
well as from electron-positron pairs from photon conversions. The latter could severely affect the distributions of R at small Q.
\end{itemize}
Fig. \ref{fig:mcgen_pions} (a) shows the simulated Q distributions for reconstructed  particles associated to
generated primary pion pairs for like and unlike-charge distributions.  One can notice that the unlike-sign pair distribution exhibits
 strong enhancements  around $Q \approx 0.35$ GeV and $Q \approx 0.7$ GeV due to $K^0$ and $\rho$ decays, respectively.  For
this reason the Q intervals $0.3 \leqslant Q \leqslant 0.45$ and $0.6 \leqslant Q \leqslant 0.825$ GeV have been excluded from the analysis.  The ratio like/unlike pairs,
eq. \ref{eq:like/unlike}, is also shown  in fig. \ref{fig:mcgen_pions}: as expected,
no structure is observed at low Q.
Adding kaons and protons we obtain the $Q$
distribution shown in fig. \ref{fig:mcgen_all} (a).  The $Q$ variable was calculated
assigning the pion mass to all particles.  No structure is visible,
only a global shift towards higher values.  Fig. \ref{fig:mcgen_all} (b),
shows the effect of adding all other particles including the ``not recognized'' ones in the
sample: the first bin is now low, demonstrating that the denominator of
eq. \ref{eq:like/unlike} contains a sizeable contribution at very low values of $Q$ due to $e^{+}e^{-}$ pairs
from photon conversions.  For this reason the data at $Q\leqslant 0.04$ GeV
have been
excluded from the fit used to extract the BEC parameters. 

We observe that the unlike-sign pair distribution as reference sample has the essential property of reproducing faithfully the non-BEC distribution of like-sign pairs (the ratio is flat). However, it is dangerously affected by  meson resonances and by  electron-positron pairs from photon conversions.
In particular the
conversions give a major contribution to our systematic errors which will be estimated in section 6.

From this study we conclude that contaminations from particles other than electrons
produce only a variation of the overall normalization of the
distributions and no distortion of its shape. Therefore they do not affect the measurement of the radius of the emitting source while some effects could be induced on the chaoticity parameter. The observation, described in section 5, that the results obtained for the ($--$)
sample are very similar to
those obtained for the (+ +) pairs, although the two samples are affected by different contaminations, demonstrates that these latter
are not a critical issue.

\section{Results}
This section presents the results on the chaoticity parameter $\lambda$ and the source
radius $R$ obtained  following the Goldhaber, KP and ($Q_{\|}$, $Q_{\bot}$) parametrizations.
\subsection{The Goldhaber parametrization}
\begin{figure}[htb]
\begin{center}
\centerline{\epsfig{file=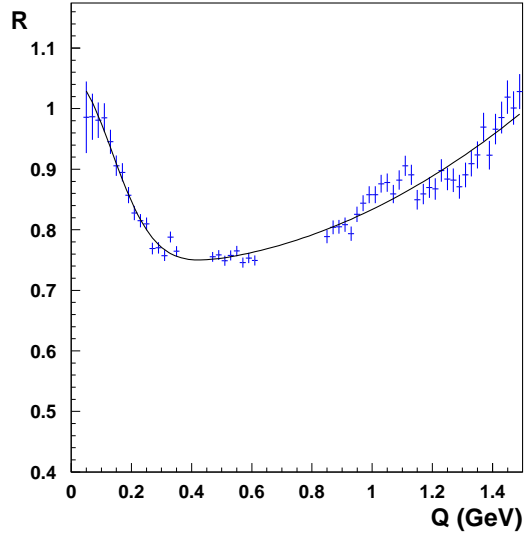,width=0.5\linewidth,angle=0}}
\caption{\small{\em R as a function of Q in the like-sign pair sample.  Superimposed
is a fit obtained using the Goldhaber parametrization (eq. \ref{eq:errecor}).}}
\label{fig:data_incl}
\end{center}
\end{figure}

 The inclusive experimental correlation R(Q) as a
function of  Q is shown in  fig. \ref{fig:data_incl}.  The empty regions in the distribution
correspond to the excluded intervals described above.
Superimposed to the data  is a fit of the form:
\begin{equation}
    R(Q) = N\big[1 + \lambda \exp(-R^2_G Q^2)\big](1 + aQ + bQ^2)
\label{eq:errecor}
\end{equation}
where $N$ is a normalization constant and the second degree polynomial is a
parametrization of the shape of the long-range correlations outside the BEC region.  The choice of the parametrization used to describe the long-range
correlations inevitably affects the results of the BEC analysis and contributes
to the systematic errors on $\lambda$ and $R_G$.  The second degree polynomial gives the best ${\chi}^2/d.o.f.$ (compared to a quadratic or linear long-range form)   of the fit and it
has been often used in other experiments. Therefore, 
we shall use it 
in this paper and we shall discuss the use of other parametrizations in the section on systematic uncertainties. 

\begin{figure}[htb]
\begin{minipage}{.5\linewidth}
\begin{center}
\centerline{\epsfig{file=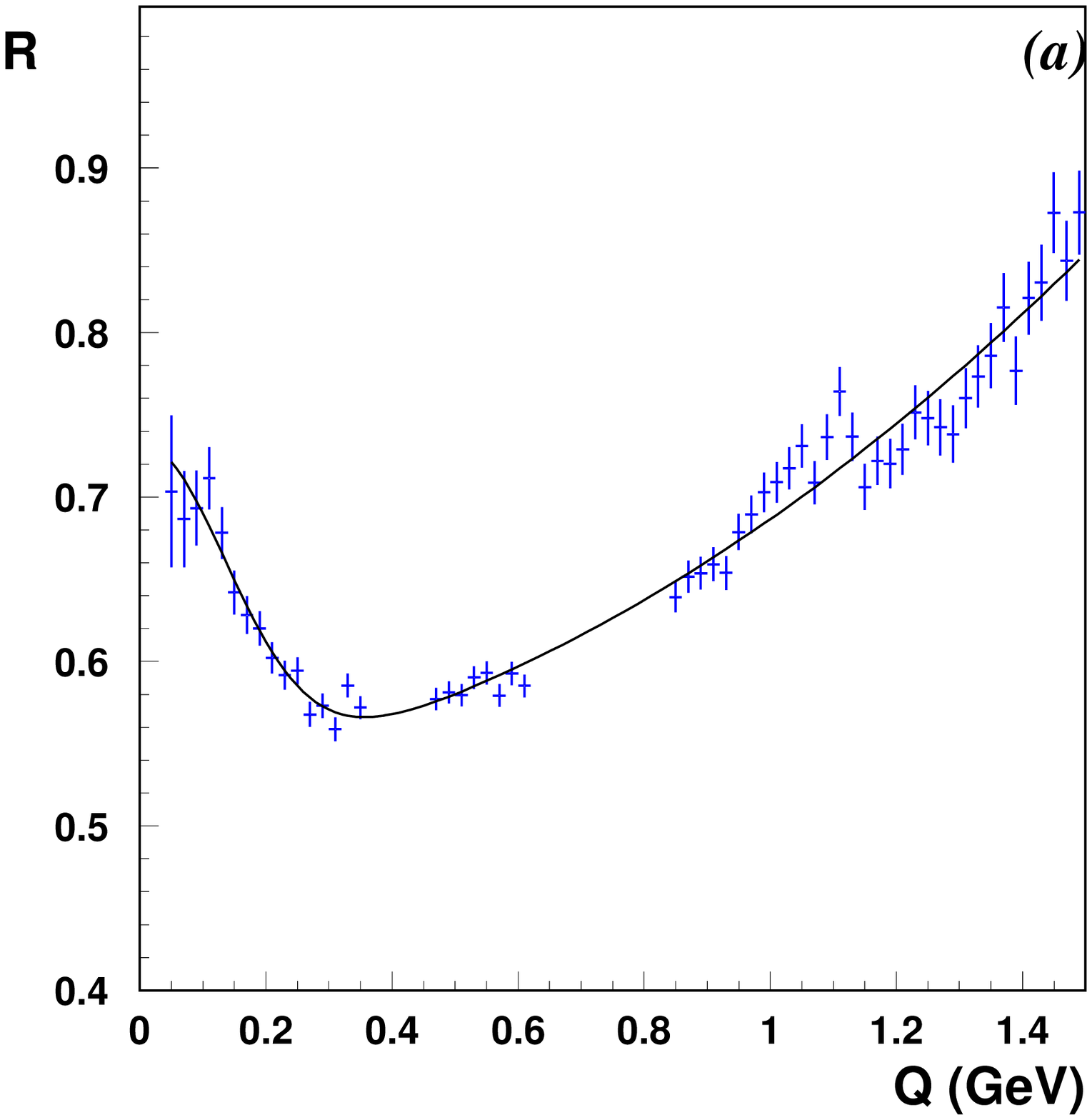,width=0.8\linewidth,angle=0}}
\end{center}
\end{minipage} \hfill
\begin{minipage}{.5\linewidth}
\begin{center}
\centerline{\epsfig{file=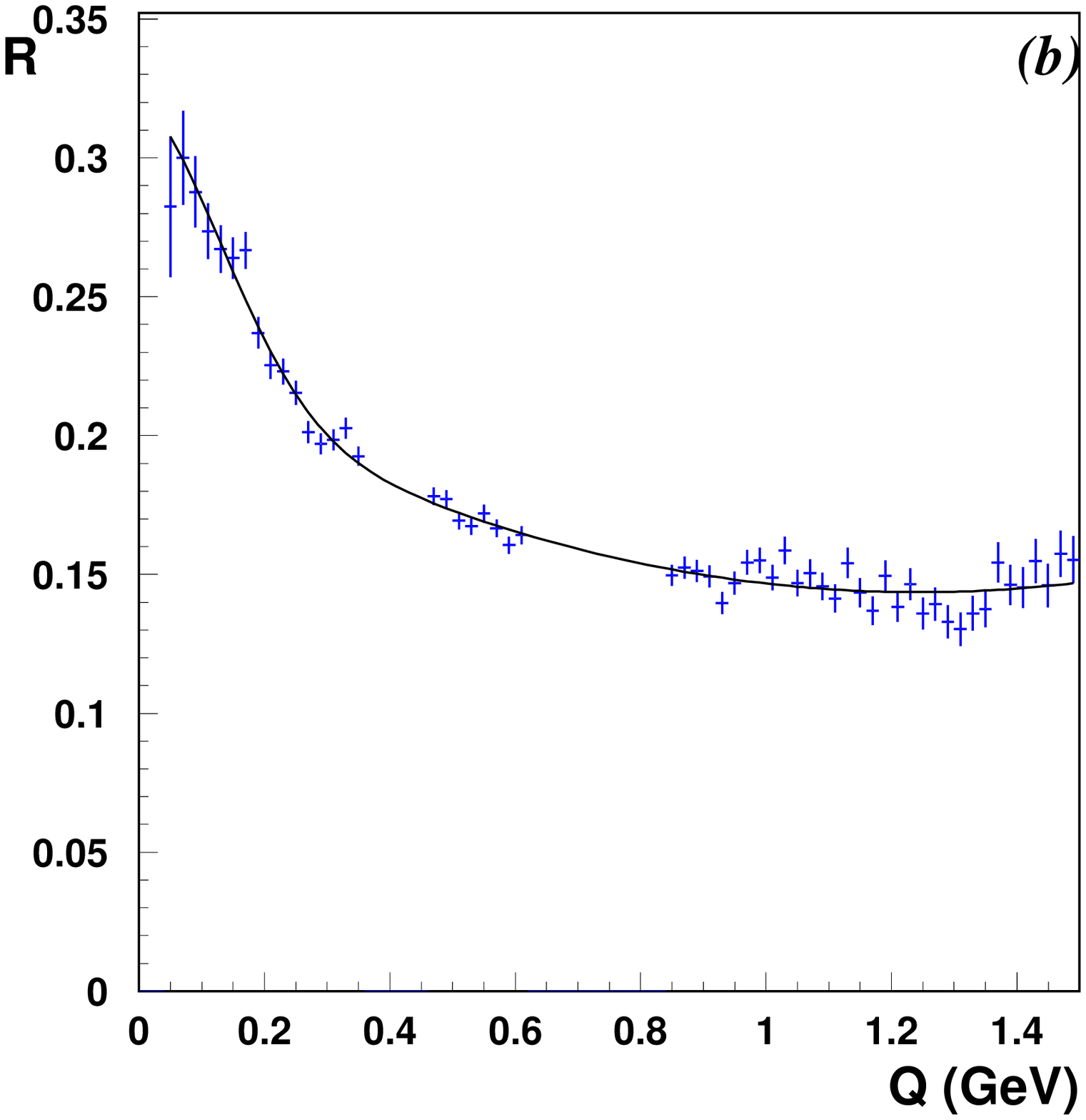,width=0.8\linewidth,angle=0}}
\end{center}
\end{minipage}
\caption{\small{\em R as a function of Q in the $(++)$ (a) and $(--)$ (b) pair
sample.  Superimposed is a fit following the Goldhaber parametrization
(eq. \ref{eq:errecor}).}}
\label{fig:data_posneg}
\end{figure}

Fig. \ref{fig:data_posneg} shows the BEC  for (+ +) pairs (a)
and ($--$) pairs (b) again with  a fit of the form of  eq. \ref{eq:errecor} superimposed.  The
long-range correlations are very different between positive and negative pairs: for positive pairs there is a steady increase of the correlation function for $Q\geqslant 0.4$ GeV, while for negative pairs the correlation function  is almost flat at large $Q$. The long-range correlations for positive pairs also determine  the behaviour of the like-sign pair  sample. However, as opposed to
the long-range correlations, 
the BEC are similar for the (++) and ($--$) samples.  Table \ref{tab:Goldhaber} summarizes the
results on  $\lambda$ and $R_{G}$ for the like, (++) and ($--$) samples. The values of $\lambda$ and $R_{G}$ of the three samples are in good agreement, in spite of the different shape of the long-range correlation region. This demonstrates that our parametrization (\ref{eq:errecor}) is robust and capable of describing correctly all three sets of data.  The BEC
parameter $\lambda$ is about 0.4 and $R_{G}$ is about 1 fm, independent of the
particle charge.
 We notice the following:
\begin{itemize}
\item The measured R distribution, in the region $Q\geqslant 0.5 $ GeV (the region of the long-range correlation) differs from the Monte Carlo simulation, which produces a flat distribution (see fig. \ref{fig:mcgen_all}). 
\item The best fit ${\chi}^2$ value is inconsistent with statistical errors alone. Most of the contribution to the
 ${\chi}^2$ comes from the region of long-range correlations ($Q\geqslant 0.8$ GeV) which is not fully accounted for by our empirical parametrization. Results from the NOMAD experiment on the production of the $f_{0}$(980) and $f_{2}$(1270) resonances in $\nu_{\mu}^{CC}$ interactions have been published   \cite{ref:resonances}. These resonances contaminate the region at $Q>0.8$ GeV and contribute to the large $\chi^{2}$ of the fit. We verified that the exclusion of the region $0.9<Q<1.3$ GeV improves the ${\chi}^2/d.o.f.$ from 1.75 to 1.5 and does not affect significantly the results. Moreover in  the BEC region  ($Q<0.2$ GeV) the quality of the fit is always good.
\end{itemize}

\begin{table}[htb]
\begin{center}
\begin{tabular}{|c|c|c|c|}
\hline
Pairs & $\lambda$ & $R_G$ (fm) & ${\chi}^2/d.o.f.$ \\
\hline
like & $0.40\pm 0.03$ & $1.01\pm 0.05$ & 90/52 \\
(+ +)  & $0.38\pm 0.04$ & $1.03\pm 0.07$ & 80/52 \\
($--$)  & $0.43\pm 0.04$ & $0.96\pm 0.06$ & 75/52 \\
\hline
\end{tabular}
\caption{\small{\em Chaoticity $\lambda$ and Goldhaber radius $R_G$.  Errors are
statistical only.}}
\label{tab:Goldhaber}
\end{center}
\end{table}
\subsection{The Kopylov-Podgoretskii parametrization}
\begin{figure}[!ht]
\begin{minipage}{.5\linewidth}
\begin{center}
\centerline{\epsfig{file=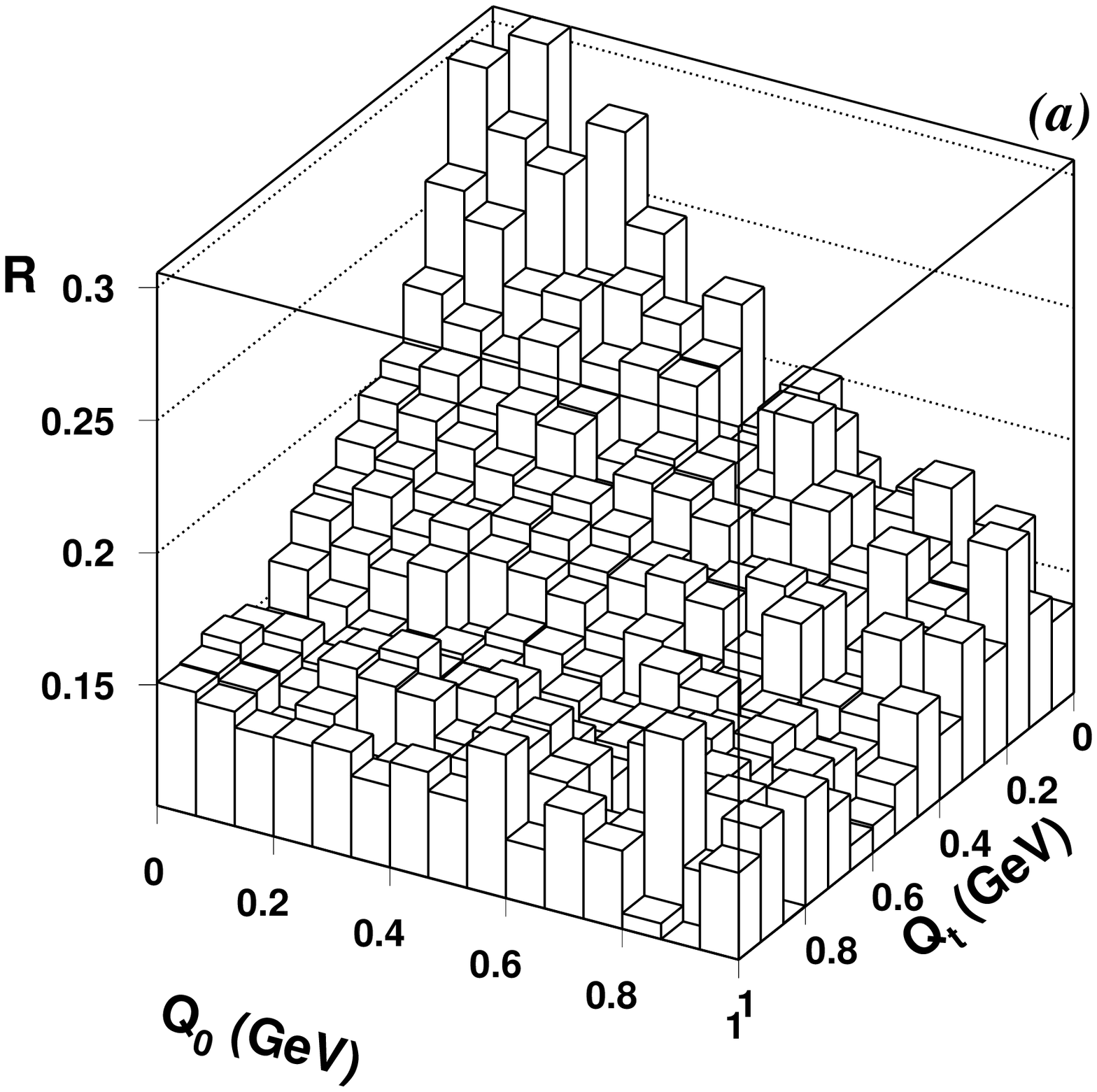,width=0.9\linewidth,angle=0}}
\end{center}
\end{minipage}
\begin{minipage}{.5\linewidth}
\begin{center}
\centerline{\epsfig{file=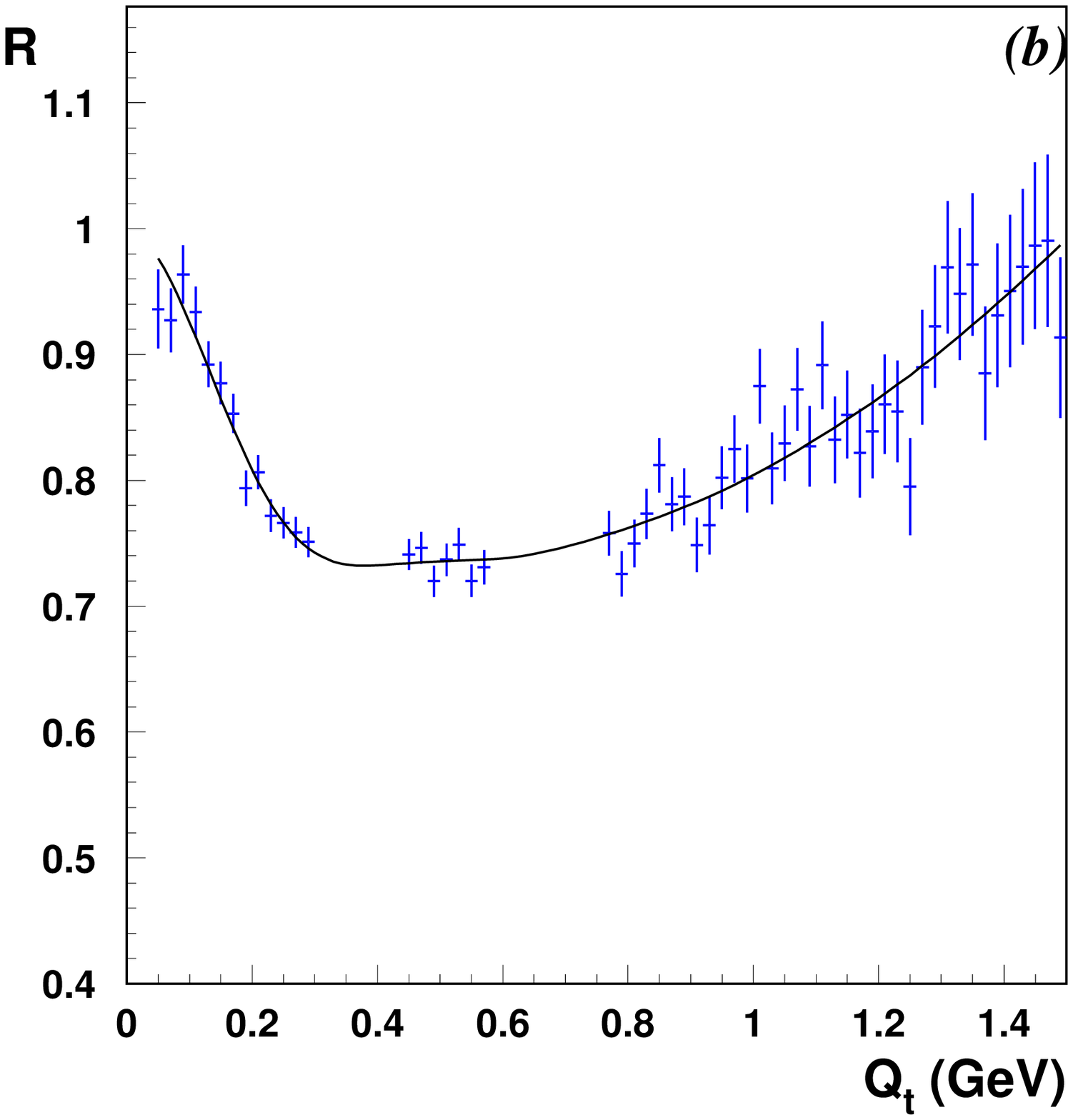,width=0.9\linewidth,angle=0}}
\end{center}
\end{minipage} \hfill
\caption{\small{\em (a), R as a function of  the  KP variables $Q_t$ and $Q_0$ for ($--$) pairs. (b), R as a function of $Q_t$ for $Q_0 \leqslant 0.2$
GeV.  Superimposed is a fit using eq. \ref{eq:kopylov1}.}}
\label{fig:Kopy_qt}
\end{figure}

 The dependence of R  on  $Q_t$ and $Q_0$  is shown in fig. \ref{fig:Kopy_qt} (a) for ($--$) pairs. A  peak at $Q_t$ and $Q_0$
$\approx 0$ is visible with a width of $\approx 0.2$ GeV in both variables.

To fit the two dimensional structure of BEC it  is convenient to use a one dimensional $Q_t$ parametrization derived from
eq. \ref{eq:kopylov2} (see \cite{ref:BEBC}, \cite{ref:AFS}) by restricting the allowed energy
difference to $Q_0 \leqslant Q_{max}$. Then, under the hypothesis that $(Q_{max}\tau)^{2}\ll 1$:
\begin{equation}
    C(Q_t) = N_{KP} \big[1 + \lambda_{KP} [2J_1(R_{KP} Q_t)/(R_{KP} Q_t)]^2\big](1 + a Q_t + b Q_t^2)
\label{eq:kopylov1}
\end{equation}
Here too a polynomial form is used to parametrize the long-range correlation.
The $Q_t$ distribution for $Q_0 \leqslant 0.2$ GeV is shown in fig. \ref{fig:Kopy_qt} (b).  Notice that also in this case we remove from the fit the   two regions where the presence of $K^0$ and resonance contributions affects the  $Q_t$ variable. 
\begin{table}[htb]
\begin{center}
\begin{tabular}{|c|c|c|c|}
\hline
Pairs & $\lambda_{KP}$ & $R_{KP}$ (fm) & ${\chi}^2/d.o.f.$ \\
\hline
like   & $0.29\pm 0.06$ & $2.07\pm 0.04$ & 51/52 \\
(+ +)  & $0.28\pm 0.04$ & $2.13\pm 0.04$ & 56/52 \\
($--$)  & $0.32\pm 0.06$ & $2.01\pm 0.04$ & 38/52 \\
\hline
\end{tabular}
\caption{\small{\em Chaoticity parameter $\lambda_{KP}$ and radius $R_{KP}$.  Errors are
statistical only.}}
\label{tab:koptab}
\end{center}
\end{table}

  The result of a fit using the
parametrization given in eq. \ref{eq:kopylov1} yields $\lambda_{KP} = 0.29\pm 0.06$
and $R_{KP} = 2.07\pm 0.04$ fm, in agreement with the expected relation $R_{KP} \approx 2R_G$. Again the results obtained using the (++) and ($--$) samples are consistent with each other.
\subsection{Longitudinal and transverse shapes}
\begin{figure}[ht!]
\begin{minipage}{.5\linewidth}
\begin{center}
\centerline{\epsfig{file=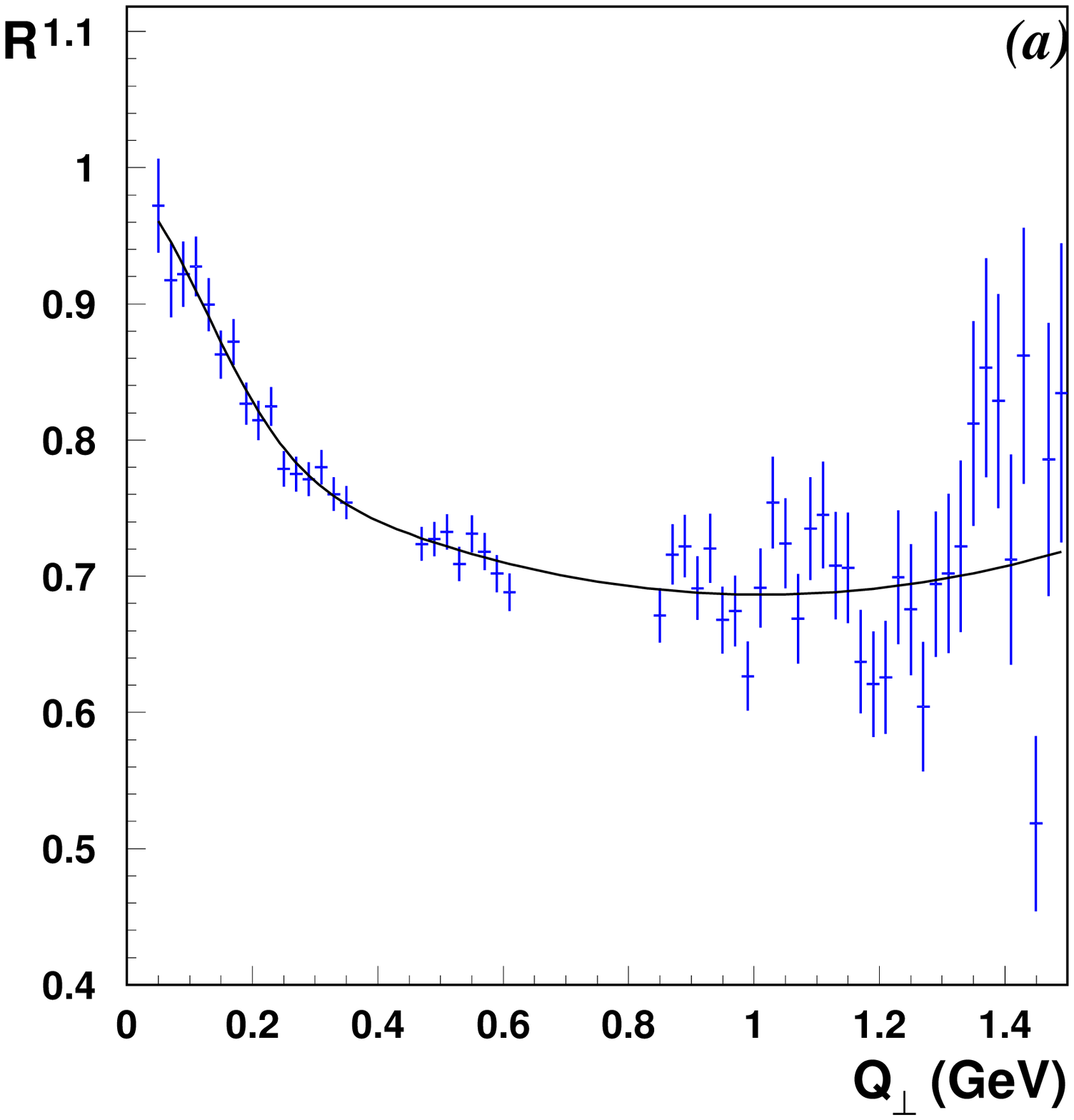,width=0.9\linewidth,angle=0}}
\end{center}
\end{minipage} \hfill
\begin{minipage}{.5\linewidth}
\begin{center}
\centerline{\epsfig{file=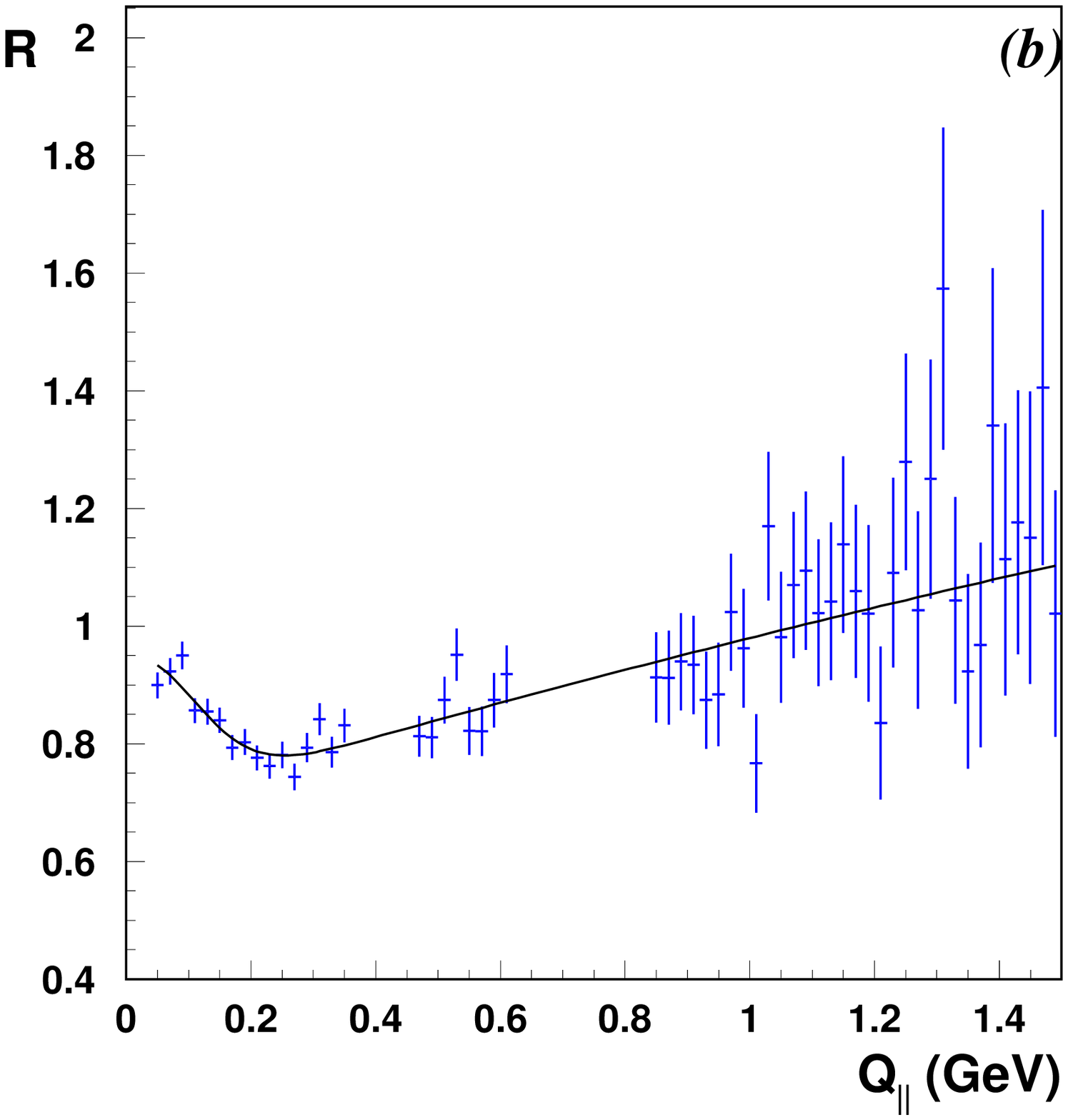,width=0.9\linewidth,angle=0}}
\end{center}
\end{minipage}
\caption{\small{\em (a), R vs $Q_{\bot}$ for
$Q_{\|}\leqslant 0.2$ GeV. (b), R vs $Q_{\|}$ for $Q_{\bot}\leqslant 0.2$ GeV
.  Superimposed  is a fit using  eq. \ref{eq:LCMSeq} multiplied
by a second degree polynomial.}}
\label{fig:qtql}
\end{figure}

A possible deviation of the pion emitting source from a spherical shape in its rest frame can be
investigated in the LCMS system using the  variables $Q_{\bot}$
and $Q_{\|}$.  The BEC behaviour is studied separately for the two variables
by requiring $Q_{\|} \leqslant 0.2$ GeV for the $Q_{\bot}$ distribution and, conversely,
$Q_{\bot} \leqslant 0.2$ GeV for the $Q_{\|}$ distribution.  These distributions are shown
in Fig. \ref{fig:qtql}. Again the regions where the presence of $K^0$'s and  resonances affects  $Q_{\|}$ and $Q_{\bot}$ have been removed from the fit. The fit is performed with a parametrization as in eq. \ref{eq:LCMSeq},
 multiplied by a second degree polynomial to reproduce the long-range
correlations. The fitted values for the BEC parameters are shown in 
tables \ref{tab:qttab} and \ref{tab:qltab} together with the results obtained from $(++)$ and $(--)$ pairs separately. Our measurements confirm
the LEP results \cite{ref:DELPHI} that in the LCMS reference frame the longitudinal
size of the pion source is 30-40\% larger than the transverse one.
\begin{table}[htb]
\begin{center}
\begin{tabular}{|c|c|c|}
\hline
Pairs &  $R_{\bot}$ (fm) & ${\chi}^2/d.o.f.$ \\
\hline
like &  $0.98\pm 0.10$ & 71/52 \\
(+ +)  &  $1.04\pm 0.12$ & 63/52 \\
($--$)  &  $0.81\pm 0.15$ & 50/52 \\
\hline
\end{tabular}
\caption{\small{\em The LCMS variable $R_{\bot}$ for
$Q_{\|}\leqslant 0.2$ GeV.  Errors are statistical only.}}
\label{tab:qttab}
\end{center}
\end{table}
\begin{table}[htb]
\begin{center}
\begin{tabular}{|c|c|c|}
\hline
Pairs &  $R_{||}$ (fm) & ${\chi}^2/d.o.f.$ \\
\hline
like &  $1.32\pm 0.14$ & 54/52 \\
(+ +)  & $1.39\pm 0.24$ & 50/52 \\
($--$)  &  $1.15\pm 0.12$ & 64/52 \\
\hline
\end{tabular}
\caption{\small{\em The LCMS variable $R_{||}$ for
$Q_{\bot}\leqslant 0.2$ GeV.  Errors are statistical only.}}
\label{tab:qltab}
\end{center}
\end{table}
\subsection{BEC dependence on the rapidity of the pair}
\begin{figure}[htb]
\begin{center}
\centerline{\epsfig{file=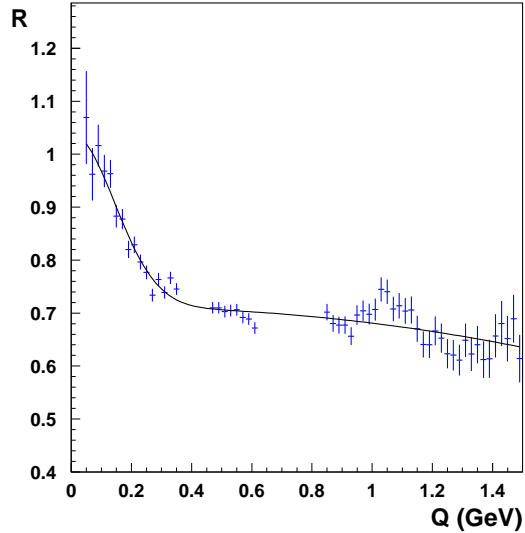,width=0.5\linewidth,angle=0}}
\caption{\small{\em R vs Q for positive
rapidity sign pairs.  Superimposed is a fit using eq. \ref{eq:errecor}.}}
\label{fig:rap_pos}
\end{center}
\end{figure}

\begin{figure}[htb]
\begin{minipage}{.5\linewidth}
\begin{center}
\centerline{\epsfig{file=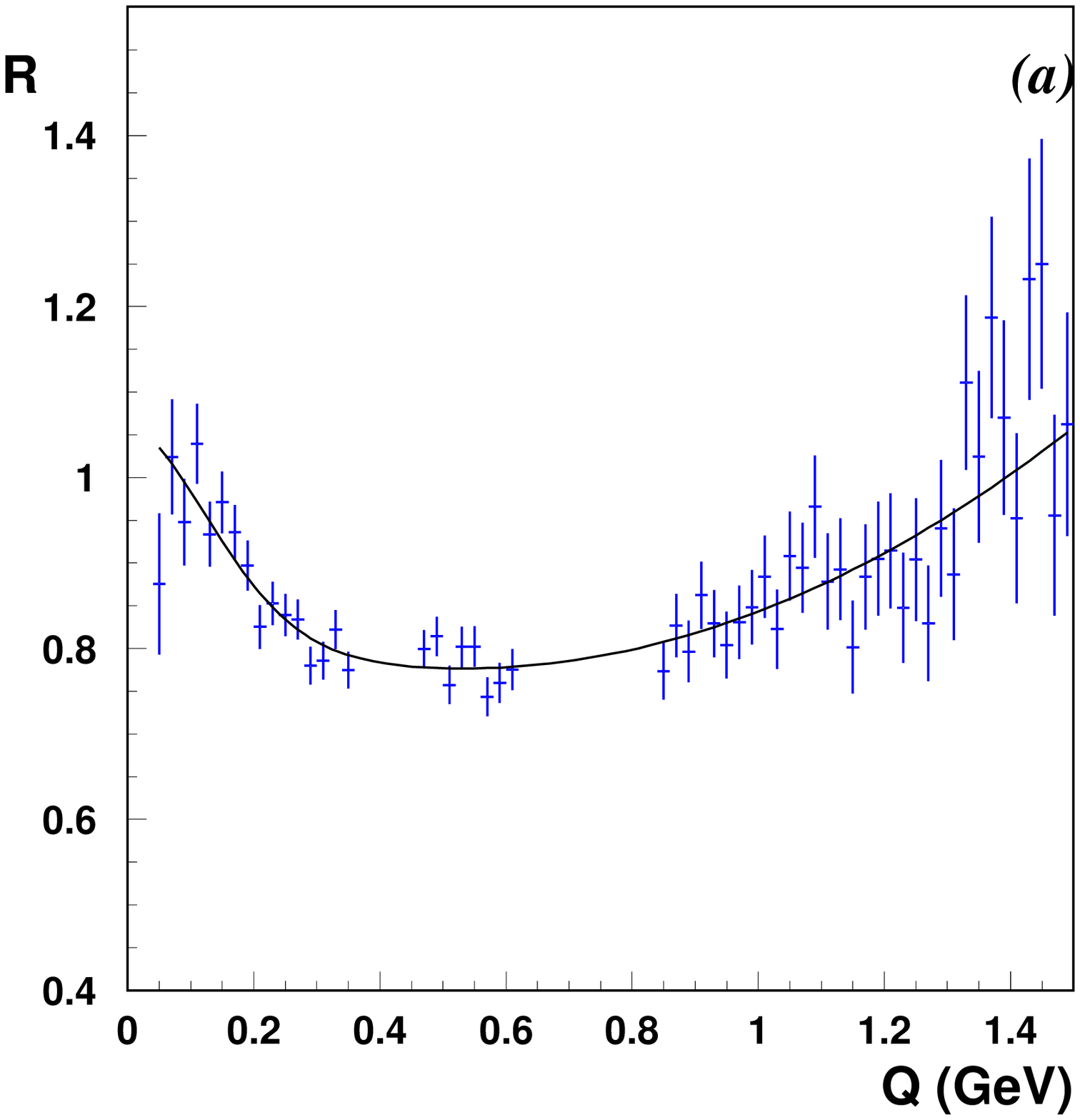,width=0.9\linewidth,angle=0}}
\end{center}
\end{minipage} \hfill
\begin{minipage}{.5\linewidth}
\begin{center}
\centerline{\epsfig{file=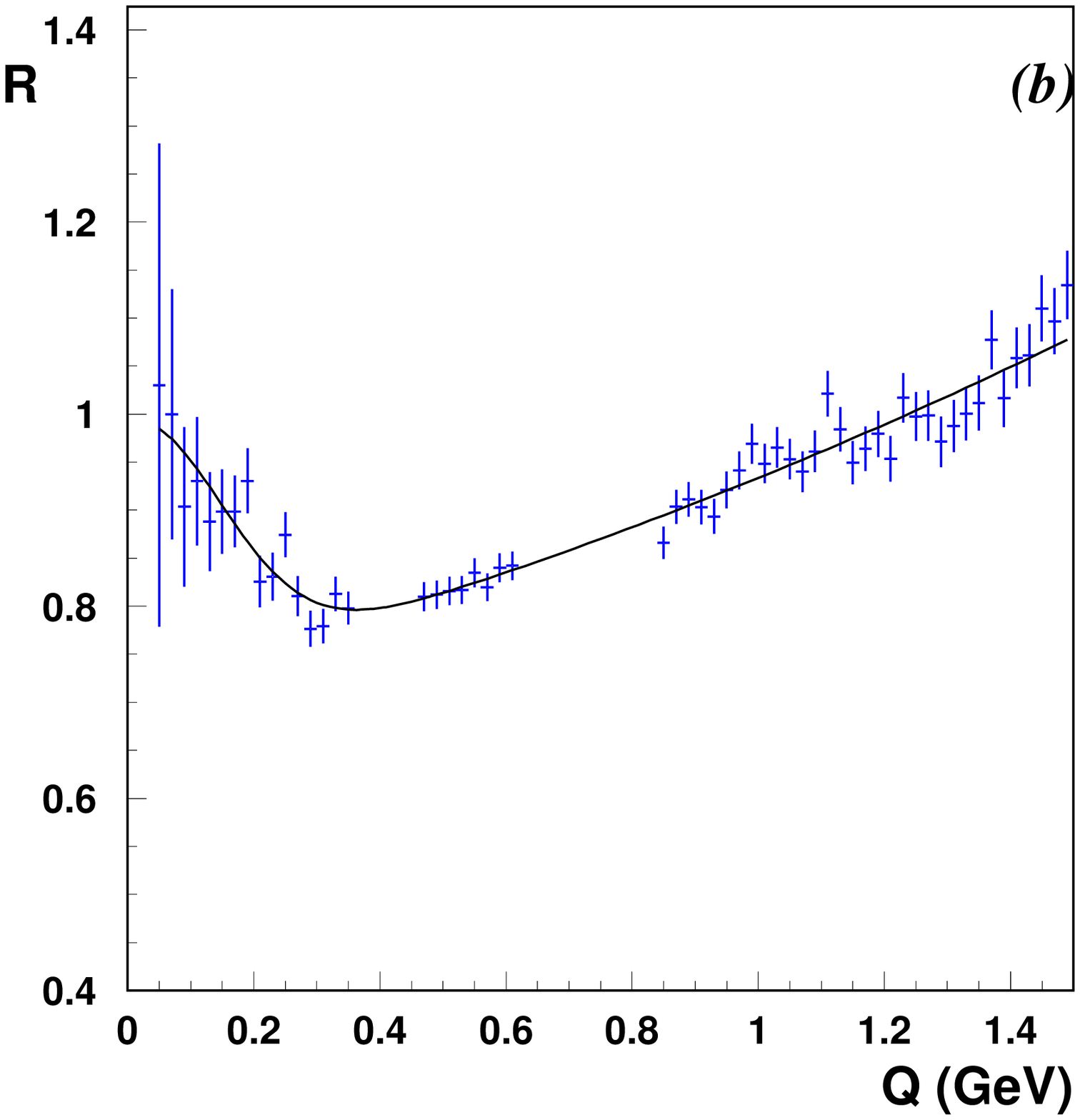,width=0.9\linewidth,angle=0}}
\end{center}
\end{minipage}
\caption{\small{\em R vs Q for negative (a) and
opposite (b) rapidity sign pairs.  Superimposed are fits using eq. \ref{eq:errecor}.}}
\label{fig:rap_neg_opp}
\end{figure}

Deep inelastic CC neutrino interactions involve a $d$ quark in 
the target nucleon leaving as spectators the remaining quarks. We naively expect, therefore,
 two distinct sources of secondary hadrons: the single struck $d$ quark
and the spectators. At high energy the two contributions should be fairly well separated in the c.m. frame of the hadronic jet as the particles coming from the fragmentation of the struck quark should have positive rapidities  while those produced by the spectator quarks should have negative rapidities.  To investigate
possible differences between the two pion sources we studied the BEC distributions,
using the Goldhaber parameter Q, for pairs of particles  of equal rapidity sign
and also for pairs of particles  of opposite rapidity sign in the rest frame of the hadronic jet. The data are shown
in fig. \ref{fig:rap_pos} and \ref{fig:rap_neg_opp}.
\begin{table}[htb]
\begin{center}
\begin{tabular}{|c|c|c|c|}
\hline
Rapidity & $\lambda$ & $R_G$ (fm) & ${\chi}^2/d.o.f.$\\
\hline
Inclusive & $0.40\pm 0.03$ & $1.01\pm 0.05$ & 90/52 \\
Positive  & $0.47\pm 0.04$ & $0.98\pm 0.07$ & 84/52 \\
Negative  & $0.42\pm 0.09$ & $1.03\pm 0.17$ & 53/52 \\
Opposite  & $0.37\pm 0.06$ & $0.98\pm 0.08$ & 52/52 \\
\hline
\end{tabular}
\caption{\small{\em Rapidity dependence of the chaoticity $\lambda$ and the Goldhaber radius $R_G$.  Errors
are statistical only.}}
\label{tab:rapidity}
\end{center}
\end{table}
Table \ref{tab:rapidity} summarizes the results obtained on  $\lambda$ and $R_G$ for the various
rapidity configurations. The source radius $R_G$ shows no differences 
for particles emitted at different rapidities, demonstrating that the typical hadronization
scale is much longer than  the interaction radius, resulting in a unique  hadron source,  independent of the detail of the quark interactions.
\subsection{BEC dependence on event  charged multiplicity and hadronic energy W }
\begin{figure}[htb]
\begin{minipage}{.5\linewidth}
\begin{center}
\centerline{\epsfig{file=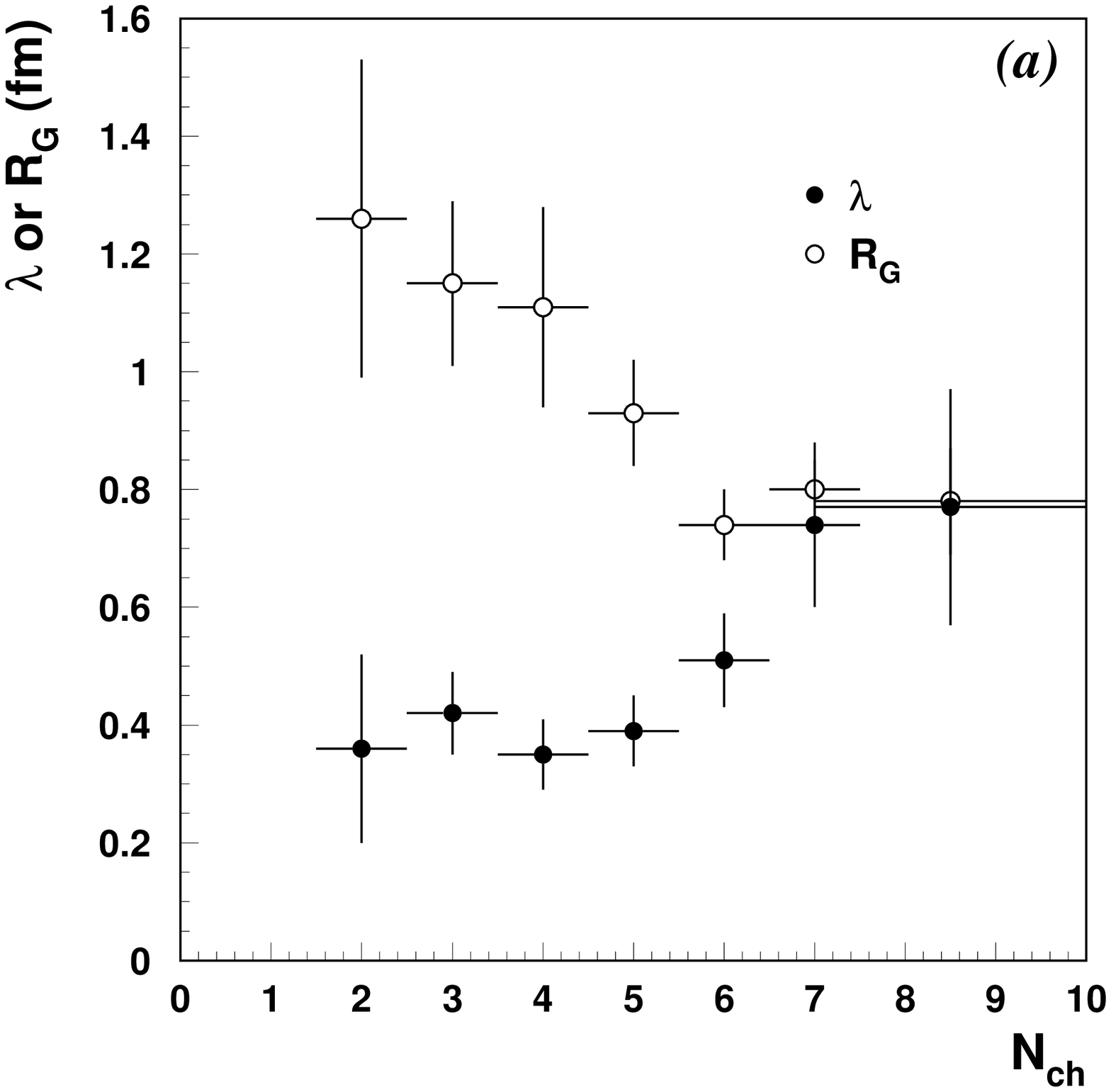,width=0.9\linewidth,angle=0}}
\end{center}
\end{minipage}
\begin{minipage}{.5\linewidth}
\begin{center}
\centerline{\epsfig{file=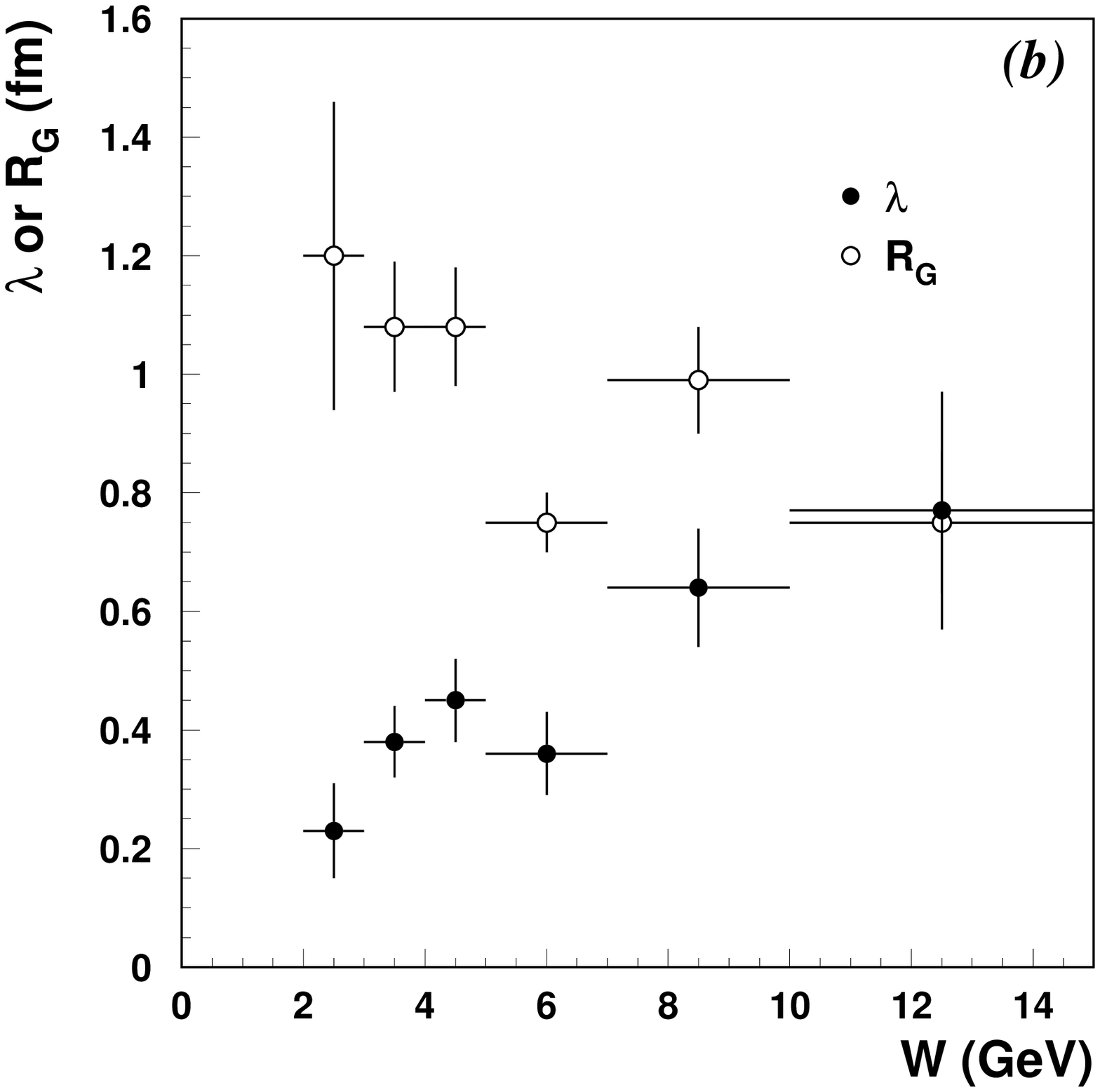,width=0.9\linewidth,angle=0}}
\end{center}
\end{minipage} \hfill
\caption{\small{\em Dependence of the chaoticity parameter $\lambda$ and Goldhaber radius $R_G$ on the event charged multiplicity $N_{ch}$ (a) and on the hadronic energy $W$ (b).}}
\label{fig:nch_w}
\end{figure}

The large number of events collected by NOMAD allows the study of BEC effects in
different final state configurations.  In particular it is interesting to
verify the observations in hadronic  \cite{ref:breakstone}
and  $e^{+}e^{-}$ interactions \cite{ref:OPAL_mult} that
the Goldhaber radius increases with the event charged multiplicity
$N_{ch}$.  Fig. \ref{fig:nch_w} (a) shows the Goldhaber radius $R_G$ and the chaoticity
parameter $\lambda$ for seven different $N_{ch}$ values. We see that there is here an indication for a decrease of $R_G$ with $N_{ch}$. One should notice that the rise of the emission radius $R_G$ with $N_{ch}$ at LEP is only visible at very high multiplicities  ($N_{ch}>10$) which are not accessible to this experiment. The chaoticity parameter $\lambda$  appears to increase   with $N_{ch}$.

A similar conclusion can be drawn when studying the BEC effects as a function of the
hadronic energy $W$: fig. \ref{fig:nch_w} (b) presents $R_G$ and $\lambda$ for six $W$ intervals. $R_{G}$  decreases with $W$ whereas $\lambda$ increases.

\section{Systematic errors}
We focus  the discussion of systematic errors to the inclusive BEC
study using the Goldhaber parametrization.

Coulomb interactions between particles which affect like-sign and unlike-sign
pairs in opposite ways, can alter the correlations.  This effect changes the
two pion cross section by the Gamow factor \cite{ref:coulomb}, a
significant correction only at very small values of $Q$.  We checked that the effect
enhances both $\lambda$ and  $R_G$ by only a few percent therefore we decided not to apply it.

We identify three sources of systematic errors in our results:
\begin{itemize}
\item The uncertainty in the background contribution underneath the BEC peak at
$Q \leqslant 0.2$ GeV.  In the  NOMAD experiment 
BEC effects could be affected by an insufficient rejection of  $e^{+}e^{-}$ pairs from
photon conversions contaminating the reference sample.
\item  The cuts applied. 
\item  The track reconstruction efficiency. We verified
that this effect does not produce sizeable effects on the Goldhaber parameters, decreasing the track reconstruction efficiency by 10\%. This was done by
removing tracks at random from the sample before calculating the  BEC.
\end{itemize}

\subsection{Systematic errors from the  $e^{+}e^{-}$  background}
As already noted in section 4.1, BEC could be altered by the presence of
background of $e^{+}e^{-}$ pairs from photon conversions in the unlike-sign sample used as a reference. Most of this background is at low $Q$ and could seriously affect the results.
To reduce this effect the data at $Q\leqslant 0.04$ GeV in all the previous correlation distributions
have been
excluded from the fit  used to extract the BEC parameters. The systematic uncertainty from this
cut is then estimated by enlarging the data excluded from the fit to $Q\leqslant 0.06$ GeV (i.e. the first bin of fig. \ref{fig:data_incl}). The results obtained for $\lambda$ and $R_G$ are shown in table \ref{tab:Goldhaber_cut}. By comparing  table \ref{tab:Goldhaber_cut} to  table \ref{tab:Goldhaber} we see that our results are insensitive to a variation of the  lowest accepted  $Q$ bin.
A larger sensitivity of $\lambda$ and $R_G$ is found when varying the track quality cuts (see next paragraph).
\begin{sloppypar}
\end{sloppypar}
\begin{table}[htb]
\begin{center}
\begin{tabular}{|c|c|c|c|}
\hline
Pairs & $\lambda$& $R_G$ (fm) & ${\chi}^2/d.o.f.$ \\
\hline
like & $0.40\pm 0.03$ &$1.02\pm 0.05$ & 89/51 \\
(+ +)  & $0.39\pm 0.03$ &$1.04\pm 0.06$ & 80/51 \\
($--$) & $0.44\pm 0.05$ &$0.98\pm 0.08$ & 84/51 \\
\hline
\end{tabular}
\caption{\small{\em Chaoticity parameter $\lambda$ and Goldhaber radius $R_G$ obtained from a fit to
the R distribution where the data at $Q\leqslant 0.06$ GeV have been removed.}}
\label{tab:Goldhaber_cut}
\end{center}
\end{table}
\subsection{Systematic errors due to the selection cuts}
We checked the stability of our results by varying the following track selection parameters:
\begin {itemize}
\item The difference $\Delta z$ between the first point of the reconstructed track and the primary vertex
position along $z$: $\Delta z \leqslant 15$ cm;
\item the minimum track momentum: $\left|\vec{p}\right| \geqslant 100$ MeV/c;
\item the maximum acceptable momentum uncertainty $\frac{\Delta \left|\vec{p}\right|}{\left|\vec{p}\right|} \leqslant 6\%$.
\end{itemize}
We note that these selection parameters affect differently pions and electrons and, therefore, they could change the fraction of conversions included in the data.

To estimate the effect of varying these cuts on the fitted parameters
$\lambda$ and $R_G$ each cut was modified and  the relevant  BEC distribution
was again  fitted. The results are shown in table \ref{tab:cutsys}.
The interval chosen for $\Delta z$ corresponds to the thickness of one DC chamber. 
The variations of the momentum and $\frac{\Delta \left|\vec{p}\right|}{\left|\vec{p}\right|}$ cut positions in table \ref{tab:cutsys} reflect the uncertainty in their choice for an optimum separation of the electron and the pion populations. A similar procedure was adopted also
for ($--$) and (+ +) pairs separately.
\begin{table}[htb]
\begin{center}
\begin{tabular}{|c|c|c|}
\hline
cut & $\lambda$ & $R_G$ (fm) \\
\hline
all cuts                           & $0.40\pm 0.03$ & $1.01\pm 0.05$ \\
$\Delta z \leqslant 10$ cm         & $0.40\pm 0.04$ & $1.10\pm 0.09$ \\
 $\Delta z \leqslant 20$ cm         & $0.37\pm 0.03$ & $0.95\pm 0.04$ \\
 $\frac{\Delta\left|\vec{p}\right|}{\left|\vec{p}\right|} \leqslant 8\%$ & $0.35\pm 0.03$ & $0.99\pm 0.04$ \\
 $\left|\vec{p}\right| \geqslant 150$ MeV/c & $0.40\pm 0.03$ & $1.01\pm 0.05$ \\
\hline
\end{tabular}
\caption{\small{\em Chaoticity parameter $\lambda$ and Goldhaber radius $R_G$ obtained for like pairs and for different
cut configurations.}}
\label{tab:cutsys}
\end{center}
\end{table}
The largest effect is the one induced on $R_G$ by changes in the $\Delta z$ cut and on $\lambda$ by changes in the $\frac{\Delta\left|\vec{p}\right|}{\left|\vec{p}\right|}$ cut. The systematic uncertainty due to the $\Delta z$ cut can be also estimated by extrapolating the parameters to $\Delta z=0$.  The variations  for $R_G$ and $\lambda$ amount to 15\% and to 12\% respectively.
\subsection{Effect of the long-range correlation parametrization}
The numerical values of $R_{G}$ and $\lambda$ depend on the parametrization used to describe the long-range correlations. In the literature linear, quadratic and polynomial forms have been used. This ambiguity must be taken into account when comparing results from different experiments.
Throughout this paper the long range correlations, necessary to
describe effects other than BEC, have been described by a second degree polynomial form
$(1 +a Q + b Q^2)$.  Different parametrizations are
possible and have been used in other experiments: i.e. a linear form:
$(1 + a Q)$ or a quadratic form: $(1 + b Q^2)$.  Table
\ref{tab:lr_par} shows the results of the three different choices for the long
range parametrizations when the like distribution is analyzed. We see that  the fit worsens
when using the linear parametrization and there are also significant
differences in the fit values for  $\lambda$ and $R_G$, while the quadratic parametrization reproduces almost exactly the results of the polynomial. We conclude that the linear approximation is inadequate to be used in the analysis of these data. For completeness in section 8
our results obtained with  different parametrizations of the long range effects will be compared with the data of other experiments using similar parametrizations.
\begin{table}[htb]
\begin{center}
\begin{tabular}{|c|c|c|c|}
\hline
Long range parametrization & $\lambda$ & $R_G$ (fm) & ${\chi}^2/d.o.f.$ \\
\hline
Polynomial & $0.40\pm 0.03$ & $1.01\pm 0.05$ & 90/52 \\
Linear     & $0.54\pm 0.02$ & $0.86\pm 0.03$ & 123/52 \\
Quadratic  & $0.43\pm 0.02$ & $0.95\pm 0.03$ & 93/52 \\
\hline
\end{tabular}
\caption{\small{\em Chaoticity parameter $\lambda$ and Goldhaber radius $R_G$ for three
choices of the long range correlation parametrization in the NOMAD analysis.  Errors
are only statistical.}}
\label{tab:lr_par}
\end{center}
\end{table}

\section{Final results}
Table \ref{tab:finres} summarizes our final results on $\lambda$ and $R_G$
including also the systematic errors from variations of the cuts discussed in the previous section (added in quadrature).
\begin{table}[htb]
\begin{center}
\begin{tabular}{|c|c|c|}
\hline
Pairs & $\lambda$& $R_{G}$ (fm) \\
\hline
like & $0.40\pm 0.03^{+0.01}_{-0.06}$ & $1.01\pm 0.05^{+0.09}_{-0.06}$ \\
(+ +)  & $0.38\pm 0.04^{+0.01}_{-0.05}$ & $1.03\pm 0.07^{+0.09}_{-0.07}$ \\
($--$)  & $0.43\pm 0.04^{+0.01}_{-0.04}$ & $0.96\pm 0.06^{+0.09}_{-0.06}$ \\
\hline
\end{tabular}
\caption{\small{\em Chaoticity $\lambda$ and Goldhaber radius $R_G$.  The first
error is statistical, the second one is systematic.}}
\label{tab:finres}
\end{center}
\end{table}
\section{Comparison with the results of other experiments}
\begin{figure}[compilation]
\begin{minipage}{.5\linewidth}
\begin{center}
\centerline{\epsfig{file=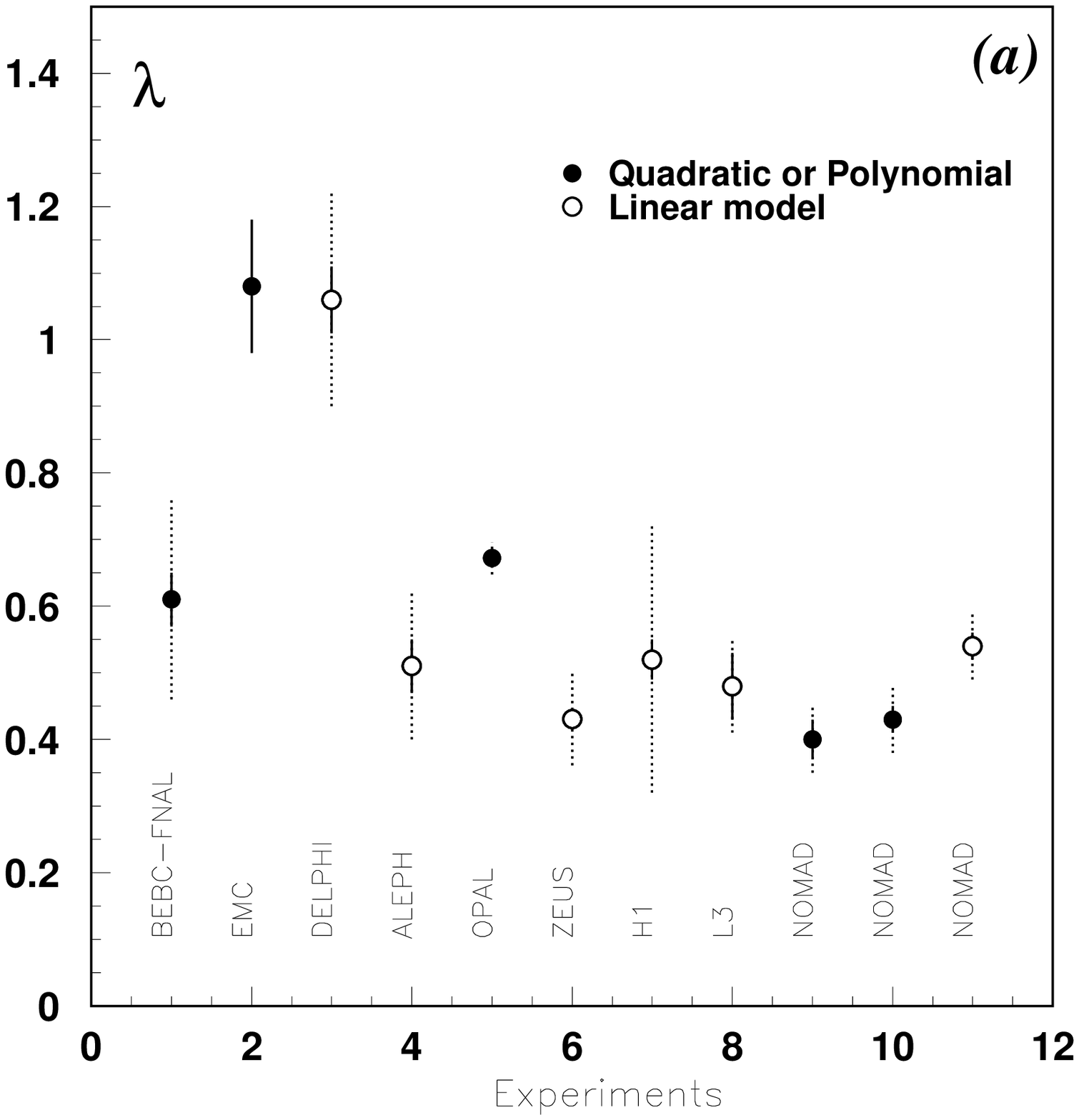,width=1.0\linewidth,angle=0}}
\end{center}
\end{minipage} \hfill 
\begin{minipage}{.5\linewidth}
\begin{center}
\centerline{\epsfig{file=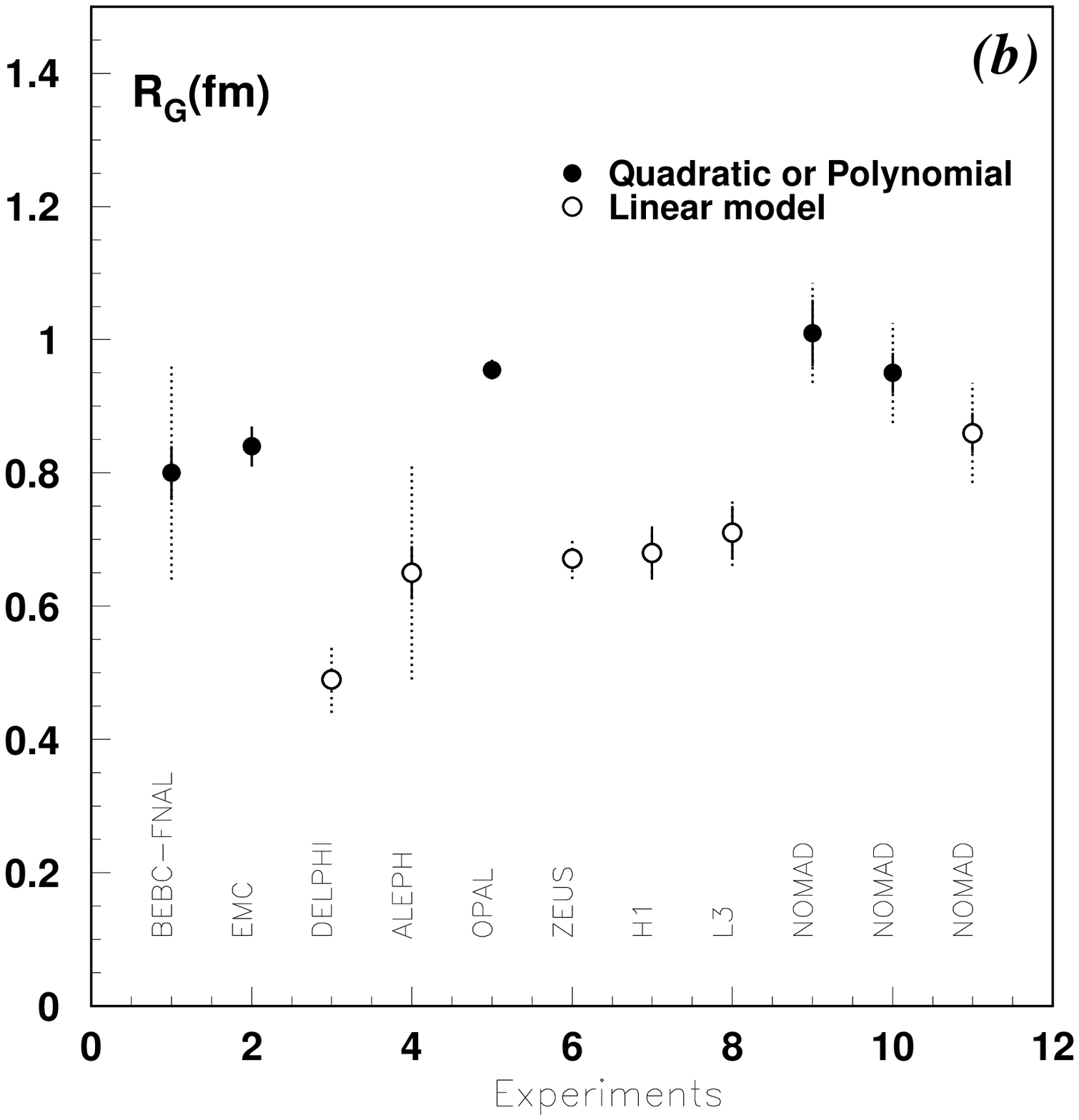,width=1.0\linewidth,angle=0}}
\end{center}
\end{minipage}
\caption{\small{\em Compilation of results obtained by various experiments for the chaoticity parameter $\lambda$ (a) and the Goldhaber radius $R_{G}$ (b).}}
\label{fig:comp}
\end{figure}

Fig. \ref{fig:comp} and table \ref{tab:others} display a compilation of some measurements of $\lambda$ and $R_{G}$ in the $\pi \pi$ channel in high statistics lepton-induced reactions: neutrino interactions \cite{ref:BEBC}, muon DIS \cite{ref:EMC}, electron-proton DIS \cite{ref:zeus}, \cite{ref:H1}, $e^{+}e^{-}$ collisions  \cite{ref:DELPHI_in},
\cite{ref:ALEPH}, \cite{ref:OPAL_mult}, \cite{ref:L3}. These experiments were performed at different  energies; they have different selection criteria and biases and also different parametrizations for the long-range correlation (see table \ref{tab:others}).

Our results agree within errors with those of the combined analysis  the BEBC and of the Fermilab neutrino data \cite{ref:BEBC}. 


The results on $\lambda$ (fig. \ref{fig:comp} (a)) shows that there are two groups of experiments which are consistent within each
group, but not between them. The first group clusters around $\lambda \approx 0.5$ and the other around $\lambda \approx 1$. The parametrization of long-range correlations does not seem to be the origin of this discrepancy. Since $\lambda$ is sensitive to the purity of the pion sample the origin of this difference could be the different pion identification criteria of the experiments.

The results on $R_{G}$ are shown in fig. \ref{fig:comp} (b): it appears that the value of $R_{G}$ computed with a linear model is systematically lower than the one computed with a quadratic or polynomial form. The two groups of data are rather well consistent within each other: the ``linear'' group cluster at $R_{G}\approx 0.6$ fm  and the ``quadratic-polynomial'' group at $R_{G}\approx 0.9$ fm.

\begin{table}[htb]
\begin{center}
\begin{tabular}{|c|c|c|c|c|c|}
\hline
Experiment & Fit type & $\langle\sqrt{s}\rangle$ &$\langle Q^2 \rangle$ &$\lambda$ & $R_G$ \\
           &          & (GeV)                    & ($\mbox{GeV}^2$)     &          & (fm)  \\
\hline
BEBC-Fermilab  & quadratic  & 10 & 10  & $0.61\pm 0.04 \pm0.15$             & $0.80\pm 0.04\pm 0.16$             \\
EMC            & quadratic  & 23 & 50  & $1.08\pm 0.1$                      & $0.84\pm 0.03$                     \\
DELPHI         & linear     & 91 &     & $1.06\pm 0.05\pm 0.16$             & $0.49\pm 0.01\pm 0.05$             \\
ALEPH          & linear     & 91 &     & $0.51\pm 0.04\pm 0.11$             & $0.65\pm 0.04\pm 0.16$             \\
ZEUS (DIS)     & linear     & 300& 400 & $0.431\pm 0.012^{+0.042}_{-0.130}$ & $0.671\pm 0.016^{+0.030}_{-0.032}$ \\
H1 (DIS)       & linear     & 300& 40  & $0.52\pm 0.03^{+0.19}_{-0.21}$     & $0.68\pm 0.04^{+0.02}_{-0.05}$     \\
L3             & linear     & 189&     & $0.48\pm 0.05\pm 0.07$             & $0.71\pm 0.04\pm 0.05$             \\
OPAL           & polynomial & 91 &     & $0.672\pm 0.013\pm 0.024$          & $0.955\pm 0.012\pm 0.015$          \\
NOMAD          & polynomial & 8  & 10  & $0.40\pm 0.03^{+0.01}_{-0.06}$      & $1.01\pm 0.05^{+0.09}_{-0.06}$     \\
NOMAD          & quadratic  & 8  & 10  & $0.43\pm 0.02$                     & $0.95\pm 0.03$                     \\
NOMAD          & linear     & 8  & 10  & $0.54\pm 0.02$                     & $0.86\pm 0.03$                     \\
\hline
\end{tabular}
\caption{\small{\em Summary of results published in previous experiments.}}
\label{tab:others}
\end{center}
\end{table}
\section{Conclusions}
The NOMAD experiment has measured BEC in charged-current neutrino interactions using different parametrizations for this effect. The general picture emerging from the data is that the size and the chaoticity of the pion source are about 1 fm and about 0.4 respectively, quite independent of the final state rapidity sign of the emitted pions. A difference of about 40\% is found between the longitudinal and transverse size of the source.  We observe a  decrease of the Goldhaber radius as a function on the  charged multiplicity and of the hadronic energy of the event. A comparison of our results with those of other experiments studying other processes than neutrino interactions  shows a fair agreement, demonstrating that the final state hadronization processes have universal features with little dependence on the type or energy of the interacting particles.
\section*{Acknowledgements}

We thank the management and staff of CERN and of all
participating institutes for their vigorous support of the experiment.
Particular thanks are due to the CERN accelerator and beam-line staff
for the magnificent performance of the neutrino beam. The following
funding agencies have contributed to this experiment:
Australian Research Council (ARC) and Department of Education, Science, and
Training (DEST), Australia;
Institut National de Physique Nucl\'eaire et Physique des Particules (IN2P3),
Commissariat \`a l'Energie Atomique (CEA), Minist\`ere de l'Education
Nationale, de l'Enseignement Sup\'erieur et de la Recherche, France;
Bundesministerium f\"ur Bildung und Forschung (BMBF, contract 05 6DO52),
Germany;
Istituto Nazionale di Fisica Nucleare (INFN), Italy;
Russian Foundation for Basic Research, 
Institute for Nuclear Research of the Russian Academy of Sciences, Russia;
Fonds National Suisse de la Recherche Scientifique, Switzerland;
Department of Energy, National Science Foundation (grant PHY-9526278),
the Sloan and the Cottrell Foundations, USA.

We also thank our secretarial staff, Jane Barney, Katherine Cross,
Joanne Hebb, Marie-Anne Huber, Jennifer Morton,
Rachel Phillips and Mabel Richtering,
and the following people who have worked with the
collaboration on the preparation and the data
collection stages of NOMAD:
M.~Anfreville, M.~Authier, G.~Barichello, A.~Beer, V.~Bonaiti, A.~Castera,
O.~Clou\'e, C.~D\'etraz, L.~Dumps, C.~Engster,
G.~Gallay, W.~Huta, E.~Lessmann,
J.~Mulon, J.P.~Pass\'e\-ri\-eux, P.~Petit\-pas, J.~Poin\-signon,
C.~Sob\-czyn\-ski, S.~Sou\-li\'e, L.~Vi\-sen\-tin, P.~Wicht.


\begin{thebibliography}{999}
\bibitem{ref:twiss} R. Hanbury-Brown and R. Q. Twiss, \phm{45} (1954) 663.
\bibitem{ref:Goldhaber} G. Goldhaber \etaltri, \prl{3} (1959) 181;\\
             G. Goldhaber \etaltri, \prl{120} (1960) 300.
\bibitem{ref:weiner} R. M. Weiner, \textit{Bose-Einstein Correlations in particle and nuclear
             physics}, John Wiley \& sons pub. (1997);\\
             R. M. Weiner, \textit{Introduction to Bose-Einstein Correlations and subatomic
             interferometry}, Chichester: Wiley, (2000).
\bibitem{ref:Kopylov} G. I. Kopylov and M. I. Podgoretskii, \sov{19} (1974) 215.
\bibitem{ref:Cocconi} G. Cocconi, \plet{49} (1974) 459.             
\bibitem{ref:BEBC}Big Bubble Chamber Neutrino Collaboration, V. A. Korotkov \etaltri,\zs{C60}(1993) 37;\\
              see also WA 25 Collaboration, D. Allasia \etaltri, \zs{C37} (1988) 527.
\bibitem{ref:NOMADET} J. Altogoer \etaltri, \nim{A428} (1999) 299.
\bibitem{ref:LEPTO} G. Ingelman, \textit{LEPTO version 6.1}, TSL-ISV-92-0065 (1992);\\
             G. Ingelman \etaltri, \textit{LEPTO version 6.5}, Comp. Phys. Comm. (1997) 101.
\bibitem{ref:JETSET} T. Sjostrand, LU-TP-95-20 (1995);\\
             T. Sjostrand, Comp. Phys. Comm. (1086) 39, (1987) 43.
\bibitem{ref:GEANT} GEANT: Detector description and simulation tool, CERN programming
             Library Long Writeups W5013, Geant version 3.21, (1994).
\bibitem{ref:DC} M. Anfreville \etaltri, \nim{A481} (2002) 339.
\bibitem{ref:TRD} G. Bassompierre \etaltri, \nim{A403} (1998) 363;\\
             G. Bassompierre \etaltri, \nim{A411} (1998) 63.
\bibitem{ref:ECAL} D. Autiero \etaltri, \nim{A373} (1996) 358;\\
             D. Autiero \etaltri, \nim{A387} (1997) 352;\\
             D. Autiero \etaltri, \nim{411} (1998) 285.
\bibitem{ref:theses}
             R. Zei, Tesi di Laurea Specialistica, Universit\`a di Pisa (2003),\\
             \textit{http://www.pi.infn.it/atlas/documenti/note/tesi zei.ps};\\
             R. C. Challis, PhD thesis, University of Melbourne (2002),\\
             \textit{http://epp.ph.unimelb.edu.au/epp/epp/epp\_current\_projects.html\#past\_theses}.
\bibitem{ref:veltri} NOMAD Collaboration, P. Astier \etaltri, \nphy{B609} (2001) 255.

\bibitem{ref:coulomb} M. Gyulassy \etaltri, \prev{C20} (1979) 2267.
\bibitem{ref:AFS} AFS Collaboration, T. Akesson \etaltri, \plet{B129} (1983) 269;\\
             AFS Collaboration, T. Akesson \etaltri, \plet{B187} (1987) 420;\\
             EHS/Na22 Collaboration, N. M. Agababian \etaltri, \zs{C59} (1993) 195.
\bibitem{ref:EMC} EMC Collaboration, M. Arneodo \etaltri, \zs{C32} (1986) 1.
\bibitem{ref:resonances} NOMAD Collaboration, P.Astier \etaltri, \nphy{B601} (2001) 3.
\bibitem{ref:OPAL_mult} OPAL Collaboration, G. Alexander \etaltri, \zs{C72} (1996) 389.
\bibitem{ref:zeus} M. Derrick, for the ZEUS Collaboration, Acta Physica Pol. \textbf{B33} (2002) 3281.
\bibitem{ref:DELPHI} DELPHI Collaboration, P. Abreu \etaltri, \plet{B471} (2000) 460.
\bibitem{ref:DELPHI_in} DELPHI Collaboration, P. Abreu \etaltri, \zs{C63} (1994) 17.
\bibitem{ref:ALEPH} ALEPH Collaboration, D. Decamp  \etaltri, \zs{C54} (1992) 75.
\bibitem{ref:L3} L3 Collaboration, M. Acciarri \etaltri, \plet{B493} (2000) 233.
\bibitem{ref:breakstone} A. Breakstone \etaltri, \zs{C33} (1987) 333.
\bibitem{ref:H1} H1 Collaboration, C. Adloff \etaltri, \zs{C75} (1997) 437.
\end{thebibliography}
\end{document}